\documentclass[12pt]{article}
\usepackage{amssymb}
\usepackage{myart}

\normalbaselines
\input tcilatex

\begin{document}

\title{Tubular geometry construction as a reason for new revision of the
space-time conception.}
\author{Yuri A.Rylov}
\date{Institute for Problems in Mechanics, Russian Academy of Sciences,\\
101-1, Vernadskii Ave., Moscow, 119526, Russia.\\
e-mail: rylov@ipmnet.ru\\
Web site: {$http://rsfq1.physics.sunysb.edu/\symbol{126}rylov/yrylov.htm$}\\
or mirror Web site: {$http://gasdyn-ipm.ipmnet.ru/\symbol{126}%
rylov/yrylov.htm$}}
\maketitle

\begin{abstract}
The tubular geometry (T-geometry) is a generalization of the proper
Euclidean geometry, founded on the property of $\sigma $-immanence. The
proper Euclidean geometry can be described completely in terms of the world
function $\sigma =\rho ^{2}/2$, where $\rho $ is the distance. This property
is called the $\sigma $-immanence. Supposing that any physical geometry is $%
\sigma $-immanent, one obtains the T-geometry $\mathcal{G}$, replacing the
Euclidean world function $\sigma _{\mathrm{E}}$ by\ means of $\sigma $ in
the $\sigma $-immanent presentation of the Euclidean geometry. One obtains
the T-geometry $\mathcal{G}$, described by the world function $\sigma $.
This method of the geometry construction is very simple and effective.
T-geometry has a new geometric property: nondegeneracy of geometry. The
class of homogeneous isotropic T-geometries is described by a form of a
function of one parameter. Using T-geometry as the space-time geometry one
can construct the deterministic space-time geometries with primordially
stochastic motion of free particles and geometrized particle mass. Such a
space-time geometry defined properly (with quantum constant as an attribute
of geometry) allows one to explain quantum effects as a result of the
statistical description of the stochastic particle motion (without a use of
quantum principles). Geometrization of the particle mass appears to be
connected with the restricted divisibility of the straight line segments.
The statement, that the problem of the elementary particle mass spectrum is
rather a problem of geometry, than that of dynamics, is a corollary of the
particle mass geometrization.
\end{abstract}

\section{Introduction}

The proper Euclidean geometry has been constructed by Euclid many years ago.
The Euclidean geometry was created as a logical construction, describing the
mutual disposition of geometrical objects in the space. The main object of
the Euclidean geometry is the point, i.e. the geometrical object, which has
no size. The straight is the geometrical object which is a continuous set of
points. The straight has a length, but it has no thickness. The plane is a
continuous set of parallel straights. The complicated geometrical objects
are combinations of more simple geometrical objects. The simplest
geometrical objects (points, straights, planes, etc.) are called elementary
geometrical objects (EGO). The Euclidean geometry is homogeneous and
isotropic, and all identical EGOs have identical properties in all places of
the space. Properties of elementary geometrical objects (EGO) are
postulated, and properties of more complicated geometrical objects are
obtained by logical reasonings from the axioms, describing properties of
EGOs. The distance between two points $P$ and $Q$ in the Euclidean geometry
is introduced as a number, describing the relation between the straight
intercept $PQ$ and some universal scale intercept $AB$. In the Euclidean
presentation of the Euclidean geometry the distance is a derivative
numerical quantity, which is not used at the Euclidean geometry construction.

On the one hand, the Euclidean geometry is a logical construction, on the
other hand, it is a science on the mutual disposition of geometrical
objects. Thus, there are two aspects of the Euclidean geometry. Any other
geometry is obtained as the Euclidean geometry generalization. One can
generalize the Euclidean geometry, considering it as a logical structure. In
this case we change properties of EGOs, i.e. change axioms of the Euclidean
geometry. As a result we obtain another homogeneous geometries (affine
geometry, projective geometry, symplectic geometry, etc.). In general, any
logical construction, which contains concepts of point and straight, can be
considered to be a geometry. We shall refer to such a geometry as the
mathematical geometry, because such geometries are interesting mainly for
mathematicians, which train their mathematical and logical capacities,
creating and investigating such geometries.

The geometry as a science on mutual disposition of geometrical objects is
interesting mainly for physicists, which use it, describing physical
phenomena in the space and in the space-time. Such a geometry will be
referred to as a physical geometry. The main characteristic of the physical
geometry is the distance between two arbitrary points of the space. A
generalization of the Euclidean geometry is obtained as a result of a
deformation of the Euclidean space. At such a deformation the distance
between points is changed. The identical EGOs in different places becomes to
be various and the obtained generalized geometry becomes inhomogeneous. In
such a geometry one cannot use axioms, because the identical EGOs become
various after inhomogeneous deformation. The generalization of the Euclidean
geometry cannot be produced in the same way, as it is produced in the
mathematical geometry. Well known mathematician Felix Klein \cite{K37}
believed that only the homogeneous geometry deserves to be called a
geometry. It is his opinion that the Riemannian geometry (in general,
inhomogeneous geometry) should be qualified as a Riemannian geography, or a
Riemannian topography. In other words, Felix Klein considered a geometry
mainly as a logical construction. We seem that the qualification of the
Riemannian geometry as a physical geometry is more appropriate, than the
Riemannian topography, although this point is not essential. It is much more
important that the physical geometry and the mathematical geometry are quite
different buildings, because their construction is founded on different
principles.

In this paper we shall consider only physical geometry, where the mutual
position of geometrical objects is the principal object of investigation.
The mutual position of geometrical objects can be described by distance $%
\rho $, which is given for all pairs of points $P,Q\in \Omega $, where $%
\Omega $ is the set of all points of the space. Usually instead of the
distance $\rho $ one considers the quantity $\sigma \left( P,Q\right) =\frac{%
1}{2}\rho ^{2}\left( P,Q\right) $, known as the world function \cite{S60}.
The world function is real even in the space-time geometry, where $\rho $
may be imaginary. It is very important that the world function of the
Euclidean geometry is the unique quantity, which is necessary for
description of the proper Euclidean geometry. In other words, \textit{the
proper Euclidean geometry can be described completely in terms and only in
terms of the world function }$\sigma _{\mathrm{E}}$ of the Euclidean space.
This statement is a very important theorem of the Euclidean geometry, which
can be proved in the framework of the Euclidean geometry. This theorem is a
foundation for construction of all physical geometries.

The property of a physical geometry of being described completely by means
of the world function will be referred to as the $\sigma $-immanence
property of this geometry. We formulate this important theorem on the $%
\sigma $-immanence of the Euclidean geometry below, as soon as the necessary
technique will be developed. Here we do note that the proper Euclidean
geometry can be constructed as a mathematical geometry without a reference
to the concept of the distance, or of the world function. The world function 
$\sigma _{\mathrm{E}}$ of the Euclidean space may be introduced, when the
proper Euclidean geometry has been already constructed. Thus we do not need
the world function for the proper Euclidean geometry construction.

The $\sigma $-immanence property of the Euclidean geometry was discovered
rather recently \cite{R90,R01}. It has been proved that the Euclidean
geometry can be presented in terms and only in terms of the function $\sigma
_{\mathrm{E}}$, provided the function $\sigma _{\mathrm{E}}$ satisfies a
series of constraints, written in terms of $\sigma _{\mathrm{E}}$. By
definition, \textit{any geometry is a totality of all geometric objects }$%
\mathcal{O}$\textit{\ and of all relations }$\mathcal{R}$\textit{\ between
them}. The $\sigma $-immanence of the proper Euclidean geometry means that
any geometric object $\mathcal{O}_{\mathrm{E}}$ and any relation $\mathcal{R}%
_{\mathrm{E}}$ of the Euclidean geometry $\mathcal{G}_{\mathrm{E}}$ can be
presented in terms of the Euclidean world function $\sigma _{\mathrm{E}}$ in
the form $\mathcal{O}_{\mathrm{E}}\left( \sigma _{\mathrm{E}}\right) $ and $%
\mathcal{R}_{\mathrm{E}}\left( \sigma _{\mathrm{E}}\right) $.

Let us suppose that any physical geometry $\mathcal{G}$ has the property of
the $\sigma $-immanence. Then the geometry $\mathcal{G}$ may be constructed
as a result of a deformation of the proper Euclidean geometry $\mathcal{G}_{%
\mathrm{E}}$. Indeed, the proper Euclidean geometry $\mathcal{G}_{\mathrm{E}%
} $ is the totality of geometrical objects $\mathcal{O}_{\mathrm{E}}\left(
\sigma _{\mathrm{E}}\right) $ and relations $\mathcal{R}_{\mathrm{E}}\left(
\sigma _{\mathrm{E}}\right) $. We produce the change

\begin{equation}
\sigma _{\mathrm{E}}\rightarrow \sigma ,\qquad \mathcal{O}_{\mathrm{E}%
}\left( \sigma _{\mathrm{E}}\right) \rightarrow \mathcal{O}_{\mathrm{E}%
}\left( \sigma \right) ,\qquad \mathcal{R}_{\mathrm{E}}\left( \sigma _{%
\mathrm{E}}\right) \rightarrow \mathcal{R}_{\mathrm{E}}\left( \sigma \right)
\label{h1.2}
\end{equation}
Then totality of geometrical objects $\mathcal{O}_{\mathrm{E}}\left( \sigma
\right) $, relations $\mathcal{R}_{\mathrm{E}}\left( \sigma \right) $ and
the world function $\sigma $ form the physical geometry $\mathcal{G}$.

For instance, let the geometrical object $\mathcal{O}_{\mathrm{E}}\left(
\sigma _{\mathrm{E}}\right) $ be a sphere $\mathcal{S}_{\mathrm{E}P_{0}Q}$,
passing through the point $Q$ and with the center at the point $P_{0}$. We
have in the proper Euclidean geometry 
\begin{equation}
\mathcal{O}_{\mathrm{E}}\left( \sigma _{\mathrm{E}}\right) :\qquad \mathcal{S%
}_{\mathrm{E}P_{0}Q}=\left\{ R|\sigma _{\mathrm{E}}\left( P_{0},R\right)
=\sigma _{\mathrm{E}}\left( P_{0},Q\right) \right\}  \label{h1.5}
\end{equation}
The geometrical object $\mathcal{O}_{\mathrm{E}}\left( \sigma \right) $ 
\begin{equation}
\mathcal{O}_{\mathrm{E}}\left( \sigma \right) :\qquad \mathcal{S}%
_{P_{0}Q}=\left\{ R|\sigma \left( P_{0},R\right) =\sigma \left(
P_{0},Q\right) \right\}  \label{h1.6}
\end{equation}
is the sphere $\mathcal{S}_{P_{0}Q}$ in the physical geometry $\mathcal{G}$.

Let $\mathcal{R}_{\mathrm{E}}\left( \sigma _{\mathrm{E}}\right) $ be the
scalar product $\left( \mathbf{P}_{0}\mathbf{P}_{1}.\mathbf{P}_{0}\mathbf{P}%
_{2}\right) _{\mathrm{E}}$ of two vectors $\mathbf{P}_{0}\mathbf{P}_{1}$, $%
\mathbf{P}_{0}\mathbf{P}_{2}$ in $\mathcal{G}_{\mathrm{E}}$. It can be
written in the $\sigma $-immanent form (i.e. in terms of the world function $%
\sigma _{\mathrm{E}}$) 
\begin{equation}
\mathcal{R}_{\mathrm{E}}\left( \sigma _{\mathrm{E}}\right) :\qquad \left( 
\mathbf{P}_{0}\mathbf{P}_{1}.\mathbf{P}_{0}\mathbf{P}_{2}\right) _{\mathrm{E}%
}=\sigma _{\mathrm{E}}\left( P_{0},P_{1}\right) +\sigma _{\mathrm{E}}\left(
P_{0},P_{2}\right) -\sigma _{\mathrm{E}}\left( P_{1},P_{2}\right)
\label{h1.3}
\end{equation}%
where index 'E' shows that the quantity relates to the Euclidean geometry.
It is easy to see that (\ref{h1.3}) is a corollary of the Euclidean
relations 
\begin{equation}
\left\vert \mathbf{P}_{0}\mathbf{P}_{1}\right\vert _{\mathrm{E}}^{2}=2\sigma
_{\mathrm{E}}\left( P_{0},P_{1}\right)  \label{h1.7}
\end{equation}%
\begin{equation}
\left\vert \mathbf{P}_{1}\mathbf{P}_{2}\right\vert _{\mathrm{E}%
}^{2}=\left\vert \mathbf{P}_{0}\mathbf{P}_{2}-\mathbf{P}_{0}\mathbf{P}%
_{1}\right\vert _{\mathrm{E}}^{2}=\left\vert \mathbf{P}_{0}\mathbf{P}%
_{1}\right\vert _{\mathrm{E}}^{2}+\left\vert \mathbf{P}_{0}\mathbf{P}%
_{2}\right\vert _{\mathrm{E}}^{2}-2\left( \mathbf{P}_{0}\mathbf{P}_{1}.%
\mathbf{P}_{0}\mathbf{P}_{2}\right) _{\mathrm{E}}  \label{h1.8}
\end{equation}

According to (\ref{h1.2}) in the physical geometry $\mathcal{G}$ we obtain
instead of (\ref{h1.3}) 
\begin{equation}
\mathcal{R}_{\mathrm{E}}\left( \sigma \right) :\qquad \left( \mathbf{P}_{0}%
\mathbf{P}_{1}.\mathbf{P}_{0}\mathbf{P}_{2}\right) =\sigma \left(
P_{0},P_{1}\right) +\sigma \left( P_{0},P_{2}\right) -\sigma \left(
P_{1},P_{2}\right)  \label{h1.4}
\end{equation}

Such a way of the physical geometry construction is very simple. It does not
use any logical reasonings. It is founded on the supposition that \textit{%
any physical geometry has the }$\sigma $\textit{-immanence property}. It
uses essentially the fact that the proper Euclidean geometry has been
already constructed, and all necessary logical reasonings has been already
produced in the proper Euclidean geometry.

The application of the replacement (\ref{h1.2}) to the construction of a
physical geometry will be referred to as the deformation principle. Any
change of distance $\rho $, or the world function $\sigma $ between the
points of the space $\Omega $ means a deformation of this space. We construe
the concept of deformation in a broad sense. The deformation may transform a
point into a surface and a surface into a point. The deformation may remove
some points of the Euclidean space, violating its continuity, or decreasing
its dimension. The deformation may add supplemental points to the Euclidean
space, increasing its dimension. We may interpret any $\sigma $-immanent
generalization of the Euclidean geometry as its deformation. In other words,
the deformation principle is a very general method of the generalized
geometry construction.

The physical geometry constructed on the basis of the deformation principle
will be referred to as the tubular geometry (T-geometry) \cite{R90,R01,R002}%
. Such a name is used, because in the T-geometry the straight lines have, in
general, a shape of hallow tubes, which in some T-geometries may degenerate
into one-dimensional curves. At this point the T-geometry distinguishes from
the Riemannian geometry, where the straight line (geodesic) is
one-dimensional by its construction.

Construction of a nonhomogeneous geometry on the axiomatic basis is
impossible practically, because there is a lot of different nonhomogeneous
geometries. It is very difficult to invent axiomatics for a nonhomogeneous
geometry, where identical objects have different properties in various
places. Besides, one cannot invent axiomatics for each of these geometries.
Thus, in reality there is no alternative to application of the deformation
principle at the construction of the physical geometry. The real problem
consists in the sequential application of the deformation principle. As far
as the \textit{deformation principle alone is sufficient for the
construction of the physical geometry}, one may not use additional means of
the geometry construction. At the physical geometry construction we do not
use coordinate system and other means of descriptions.

Usually a construction of the Riemannian geometry is carried out in some
coordinate system. The Riemannian geometry is obtained from the Euclidean
one by means of the deformation principle, i.e. by the change infinitesimal
Euclidean distance $dS_{\mathrm{E}}=\sqrt{g_{\mathrm{E}ik}dx^{i}dx^{k}}$ by
means of the Riemannian one $dS_{\mathrm{R}}=\sqrt{g_{\mathrm{R}%
ik}dx^{i}dx^{k}}$. Properties of the Riemannian geometry are determined by
the form of the metric tensor $g_{\mathrm{R}ik}$. But the form of the metric
tensor depends on the choice of the coordinate system. If we use different
coordinate systems, we obtain formally different description of the same
Riemannian geometry. To separate the essential part of description, which
relates to the geometry in itself, we are forced to consider description in
all possible coordinate systems. The common part of descriptions in all
coordinate systems (invariants of the group of the coordinate
transformations) forms the description of the geometry in itself.
Unfortunately, we cannot use all possible coordinate systems. Practically we
use only continuous coordinates. The number of coordinates is fixed, and
coincides with the dimension of the Riemannian geometry. We cannot solve
definitely, whether the continuity is a property of the considered geometry,
or maybe, it is a property of the coordinate description. As a result, most
geometers believe that the continuity is the inherent property of the
geometry. They admit that the discrete geometry may be constructed, but they
do not know, how to do this, because one cannot use discrete coordinates for
description of discrete geometries.

Mathematicians provide physicists with their geometrical construction, and
physicists believe that the space-time is continuous. Continuity of the
space-time cannot be tested experimentally, and the only reason of the
space-time continuity is the fact that mathematicians are able to construct
only continuous geometries, whereas they fail to construct discrete
geometries.

We have the same situation with the space dimension. Geometers consider the
dimension to be an inherent property of any geometry. They can imagine the $%
n $-dimensional Riemannian geometry, but they cannot imagine a geometry
without a dimension, or a geometry of an indefinite dimension. The reason of
these belief is the fact that the dimension of the manifold and its
continuity are the starting points of the Riemannian geometry construction,
and at this point one cannot separate the properties of the geometry from
the properties of the manifold.

In the T-geometry we deal only with the geometry in itself, because it does
not use any means of the description. As a result the T-geometry is
insensitive to continuity or discreteness of the space, as well as to its
dimension. Application of additional means of description can lead to
inconsistency and to a restriction of the list of possible physical
geometries.

Any generalization of the proper Euclidean geometry is founded on some
property of the Euclidean geometry (or its objects). This property is
conserved in all generalized geometries, whereas other properties of the
Euclidean geometry are varied. Character and properties of the obtained
generalized geometry depend essentially on the choice of the conserved
property of the basic Euclidean geometry. For instance, the Riemannian
geometry is such a generalization of the Euclidean one, where the
one-dimensionality and continuity of the Euclidean straight line are
conserved, whereas its curvature and torsion are varied. The straight line
is considered to be the principal geometric object of the Euclidean
geometry, and one supposes that such properties of the Euclidean straight
line as continuity and one-dimensionality (absence of thickness) are to be
conserved at the generalization. It means that the continuity and
one-dimensionality of the straight line are to be the principal concepts of
the generalized geometry (the Riemannian geometry). In accordance with such
a choice of the conserved geometrical object one introduces the concept of
the curve $\mathcal{L}$ as a continuous mapping of a segment of the real
axis onto the space $\Omega $%
\begin{equation}
\mathcal{L}:\qquad \left[ 0,1\right] \rightarrow \Omega  \label{g1.1}
\end{equation}
To introduce the concept of the continuity, which is a basic concept of the
generalization, one introduces the topological space, the dimension of the
space $\Omega $ and other basic concepts of the Riemannian geometry, which
are necessary for construction of the Riemannian generalization of the
Euclidean geometry.

The $\sigma $-immanence of the Euclidean geometry is a property of the whole
Euclidean geometry. Using the $\sigma $-immanence for generalization, we do
not impose any constraints on the single geometric objects of the Euclidean
geometry. As a result the $\sigma $-immanent generalization appears to be a
very powerful generalization. Besides, from the common point of view the
application of the whole geometry property for the generalization seems to
be more reasonable, than a use of the properties of a single geometric
object. Thus, using the property of the whole Euclidean geometry, the $%
\sigma $-immanent generalization seems to be more reasonable, than the
Riemannian generalization, using the properties of the Euclidean straight
line.

Now we list the most attractive features of the $\sigma $-immanent
generalization of the Euclidean geometry:

\begin{enumerate}
\item It uses for the generalization the $\sigma $-immanence, which is a
property of the Euclidean geometry as a whole (but not a property of a
single geometric object as it takes place at the Riemannian generalization).

\item The $\sigma $-immanent generalization does not use any logical
construction, and the $\sigma $-immanent generalization is automatically as
consistent, as the Euclidean geometry, whose axiomatics is used implicitly.
In particular, the T-geometry does not contain any theorems. As a result the
main problem of the T-geometry is a correct $\sigma $-immanent description
of geometrical objects and relations of the Euclidean geometry. There are
some subtleties in such a $\sigma $-immanent description, which are
discussed below.

\item The $\sigma $-immanent generalization is a very powerful
generalization. It varies practically all properties of the Euclidean
geometry, including such ones as the continuity and the parallelism
transitivity, which are conserved at the Riemannian generalization.

\item The $\sigma $-immanent generalization allows one to use the
coordinateless description and ignore the problems, connected with the
coordinate transformations as well as with the transformation of other means
of description.

\item The T-geometry may be used as the space-time geometry. In this case
the tubular character of straights explains freely the stochastic world
lines of quantum microparticles. Considering the quantum constant $\hbar $
as an attribute of the space-time geometry, one can obtain the quantum
description as the statistical description of the stochastic world lines 
\cite{R91}. Such a space-time geometry cannot be obtained in the framework
of the Riemannian generalization of the Euclidean geometry.
\end{enumerate}

In the paper we discuss the interplay between the Riemannian generalization
of the Euclidean geometry and the $\sigma $-immanent one. The fact is that
the world function $\sigma $ has been introduced at first in the Riemannian
geometry \cite{S60}. The world function $\sigma $ plays an important, but
not crucial role in the description of the Riemannian geometry, whereas in
the T-geometry the world function $\sigma $ is the only quantity, which is
necessary for its construction and description.

Let us consider two generalization of the Euclidean geometry $\mathcal{G}_{%
\mathrm{E}}$ with the same world function $\sigma $. Let one of the
generalization be the Riemannian geometry $\mathcal{G}_{\mathrm{R}}$ and the
other one be a $\sigma $-immanent generalization $\mathcal{G}_{\sigma }$. Do
the generalizations $\mathcal{G}_{\mathrm{R}}$ and $\mathcal{G}_{\sigma }$
coincide? Many important details of both generalizations coincide, but other
details of the geometrical description are different. In general, the
geometries $\mathcal{G}_{\mathrm{R}}$ and $\mathcal{G}_{\sigma }$ are
different, although both geometries $\mathcal{G}_{\mathrm{R}}$ and $\mathcal{%
G}_{\sigma }$ are described by the same world function $\sigma $. The
geometry $\mathcal{G}_{\sigma }$ is determined by the world function $\sigma 
$ uniquely without any logical constructions. It means that the $\sigma $%
-immanent generalization $\mathcal{G}_{\sigma }$ cannot be inconsistent,
because the basic geometry $\mathcal{G}_{\mathrm{E}}$ is consistent \cite%
{H30}. As to the Riemannian generalization $\mathcal{G}_{\mathrm{R}}$, it
may be inconsistent, because it uses additional means of description
(manifold, topology), which are not necessary for construction of the $%
\sigma $-immanent generalization of the Euclidean geometry. One should
investigate to what extent the additional means of description are
compatible between themselves and with the given world function $\sigma $,
which is alone sufficient for the construction of the consistent $\sigma $%
-immanent generalization $\mathcal{G}_{\sigma }$.

Application of additional means of description leads to an overdetermination
of the problem of the Euclidean geometry generalization. Some
inconsistencies of the Riemannian generalization are corollaries of this
overdetermination.

\section{Euclidean geometry in the $\protect\sigma $-immanent form}

\begin{definition}
The $\sigma $-space $V=\left\{ \sigma ,\Omega \right\} $ is the set $\Omega $
of points $P$ with the given world function $\sigma $
\begin{equation}
\sigma :\qquad \Omega \times \Omega \rightarrow \mathbb{R},\qquad \sigma
\left( P,P\right) =0,\qquad \forall P\in \Omega   \label{h1.1}
\end{equation}
\end{definition}

Let the proper Euclidean geometry be given on the set $\Omega $, and the
quantity 
\begin{equation}
\rho \left( P_{0},P_{1}\right) =\sqrt{2\sigma \left( P_{0},P_{1}\right) }%
,\qquad P_{0},P_{1}\in \Omega  \label{g2.1}
\end{equation}%
be the Euclidean distance between the points $P_{0},P_{1}$.

Let the vector $\mathbf{P}_{0}\mathbf{P}_{1}\mathbf{=}\left\{
P_{0},P_{1}\right\} $ be the ordered set of two points $P_{0}$, $P_{1}$. The
point $P_{0}$ is the origin of the vector $\mathbf{P}_{0}\mathbf{P}_{1}$,
and the point $P_{1}$ is its end. The length $\left\vert \mathbf{P}_{0}%
\mathbf{P}_{1}\right\vert $ of the vector $\mathbf{P}_{0}\mathbf{P}_{1}$ is
defined by the relation 
\begin{equation}
\left\vert \mathbf{P}_{0}\mathbf{P}_{1}\right\vert ^{2}=2\sigma \left(
P_{0},P_{1}\right)  \label{g2.2}
\end{equation}%
In the Euclidean geometry the scalar product $\left( \mathbf{P}_{0}\mathbf{P}%
_{1}.\mathbf{P}_{0}\mathbf{P}_{2}\right) $ of two vectors $\mathbf{P}_{0}%
\mathbf{P}_{1}$ and $\mathbf{P}_{0}\mathbf{P}_{2}$, having the common origin 
$P_{0}$, is expressed by the relation (\ref{h1.4}) 
\begin{equation}
\left( \mathbf{P}_{0}\mathbf{P}_{1}.\mathbf{P}_{0}\mathbf{P}_{2}\right)
=\sigma \left( P_{0},P_{1}\right) +\sigma \left( P_{1},P_{0}\right) -\sigma
\left( P_{1},P_{2}\right)  \label{g2.3}
\end{equation}

It follows from the expression (\ref{g2.3}), written for scalar products $%
\left( \mathbf{P}_{0}\mathbf{P}_{1}.\mathbf{P}_{0}\mathbf{Q}_{1}\right) $
and $\left( \mathbf{P}_{0}\mathbf{P}_{1}.\mathbf{P}_{0}\mathbf{Q}_{0}\right) 
$, and from the properties of the scalar product in the Euclidean space,
that the scalar product $\left( \mathbf{P}_{0}\mathbf{P}_{1}.\mathbf{Q}_{0}%
\mathbf{Q}_{1}\right) $ of two vectors $\mathbf{P}_{0}\mathbf{P}_{1}$ and $%
\mathbf{Q}_{0}\mathbf{Q}_{1}$ can be written in the $\sigma $-immanent form 
\begin{eqnarray}
\left( \mathbf{P}_{0}\mathbf{P}_{1}.\mathbf{Q}_{0}\mathbf{Q}_{1}\right)
&=&\left( \mathbf{P}_{0}\mathbf{P}_{1}.\mathbf{P}_{0}\mathbf{Q}_{1}\right)
-\left( \mathbf{P}_{0}\mathbf{P}_{1}.\mathbf{P}_{0}\mathbf{Q}_{0}\right) = 
\nonumber \\
&&\sigma \left( P_{0},Q_{1}\right) +\sigma \left( P_{1},Q_{0}\right) -\sigma
\left( P_{0},Q_{0}\right) -\sigma \left( P_{1},Q_{1}\right)  \label{b1.1a}
\end{eqnarray}

Let $\mathbf{P}_{0}\mathbf{P}_{1}$, $\mathbf{P}_{0}\mathbf{P}_{2}$,...$%
\mathbf{P}_{0}\mathbf{P}_{n}$ be $n$ vectors in the Euclidean space. The
necessary and sufficient condition of their linear dependence is 
\begin{equation}
F_{n}\left( \mathcal{P}^{n}\right) \equiv \det \left\vert \left\vert \left( 
\mathbf{P}_{0}\mathbf{P}_{i}.\mathbf{P}_{0}\mathbf{P}_{k}\right) \right\vert
\right\vert =0,\qquad i,k=1,2,..n,\qquad \mathcal{P}^{n}=\left\{
P_{0},P_{1},...P_{n}\right\}  \label{g2.4}
\end{equation}%
where $F_{n}\left( \mathcal{P}^{n}\right) \equiv \det \left\vert \left\vert
\left( \mathbf{P}_{0}\mathbf{P}_{i}.\mathbf{P}_{0}\mathbf{P}_{k}\right)
\right\vert \right\vert $ is the Gram's determinant, constructed of the
scalar products of vectors.

Let us formulate the theorem on the $\sigma $-immanence of the Euclidean
geometry.

\begin{theorem}
The $\sigma $-space $V=\left\{ \sigma ,\Omega \right\} $ is the $n$%
-dimensional proper Euclidean space, if and only if the world function $%
\sigma $ satisfies the following conditions, written in terms of the world
function $\sigma $.
\end{theorem}

\textit{I. Condition of symmetry: } 
\begin{equation}
\sigma \left( P,Q\right) =\sigma \left( Q,P\right) ,\qquad \forall P,Q\in
\Omega  \label{a1.4}
\end{equation}

\textit{II. Definition of the dimension: } 
\begin{equation}
\exists \mathcal{P}^{n}\equiv \left\{ P_{0},P_{1},...P_{n}\right\} \subset
\Omega ,\qquad F_{n}\left( \mathcal{P}^{n}\right) \neq 0,\qquad F_{k}\left( {%
\Omega }^{k+1}\right) =0,\qquad k>n  \label{g2.5}
\end{equation}%
\textit{where }$F_{n}\left( \mathcal{P}^{n}\right) $\textit{\ is the Gram's
determinant (\ref{g2.4}). Vectors }$\mathbf{P}_{0}\mathbf{P}_{i}$\textit{, }$%
\;i=1,2,...n$\textit{\ are basic vectors of the rectilinear coordinate
system }$K_{n}$\textit{\ with the origin at the point }$P_{0}$\textit{, and
the metric tensors }$g_{ik}\left( \mathcal{P}^{n}\right) $\textit{, }$%
g^{ik}\left( \mathcal{P}^{n}\right) $\textit{, \ }$i,k=1,2,...n$\textit{\ in 
}$K_{n}$\textit{\ are defined by the relations } 
\begin{equation}
\sum\limits_{k=1}^{k=n}g^{ik}\left( \mathcal{P}^{n}\right) g_{lk}\left( 
\mathcal{P}^{n}\right) =\delta _{l}^{i},\qquad g_{il}\left( \mathcal{P}%
^{n}\right) =\left( \mathbf{P}_{0}\mathbf{P}_{i}.\mathbf{P}_{0}\mathbf{P}%
_{l}\right) ,\qquad i,l=1,2,...n  \label{a1.5b}
\end{equation}%
\begin{equation}
F_{n}\left( \mathcal{P}^{n}\right) =\det \left\vert \left\vert g_{ik}\left( 
\mathcal{P}^{n}\right) \right\vert \right\vert \neq 0,\qquad i,k=1,2,...n
\label{g2.6}
\end{equation}

\textit{III. Linear structure of the Euclidean space: } 
\begin{equation}
\sigma \left( P,Q\right) =\frac{1}{2}\sum\limits_{i,k=1}^{i,k=n}g^{ik}\left( 
\mathcal{P}^{n}\right) \left( x_{i}\left( P\right) -x_{i}\left( Q\right)
\right) \left( x_{k}\left( P\right) -x_{k}\left( Q\right) \right) ,\qquad
\forall P,Q\in \Omega  \label{a1.5a}
\end{equation}%
\textit{where coordinates }$x_{i}\left( P\right) ,$\textit{\ }$i=1,2,...n$%
\textit{\ of the point }$P$\textit{\ are covariant coordinates of the vector 
}$\mathbf{P}_{0}\mathbf{P}$\textit{, defined by the relation } 
\begin{equation}
x_{i}\left( P\right) =\left( \mathbf{P}_{0}\mathbf{P}_{i}.\mathbf{P}_{0}%
\mathbf{P}\right) ,\qquad i=1,2,...n  \label{b12}
\end{equation}

\textit{IV: The metric tensor matrix }$g_{lk}\left( \mathcal{P}^{n}\right) $%
\textit{\ has only positive eigenvalues } 
\begin{equation}
g_{k}>0,\qquad k=1,2,...,n  \label{a15c}
\end{equation}

\textit{V. The continuity condition: the system of equations } 
\begin{equation}
\left( \mathbf{P}_{0}\mathbf{P}_{i}.\mathbf{P}_{0}\mathbf{P}\right)
=y_{i}\in \mathbb{R},\qquad i=1,2,...n  \label{b14}
\end{equation}%
\textit{considered to be equations for determination of the point }$P$%
\textit{\ as a function of coordinates }$y=\left\{ y_{i}\right\} $\textit{,\
\ }$i=1,2,...n$\textit{\ has always one and only one solution. }Conditions
II -- V contain a reference to the dimension $n$\ of the Euclidean space.

This theorem states that the proper Euclidean space has the property of the $%
\sigma $-immanence, and hence any statement $\mathcal{S}$ of the proper
Euclidean geometry can be expressed in terms and only in terms of the world
function $\sigma _{\mathrm{E}}$ of the Euclidean geometry in the form $%
\mathcal{S}\left( \sigma _{\mathrm{E}}\right) $. Producing the change $%
\sigma _{\mathrm{E}}\rightarrow \sigma $ in the statement $\mathcal{S}$, we
obtain corresponding statement $\mathcal{S}\left( \sigma \right) $ of
another T-geometry $\mathcal{G}$, described by the world function $\sigma $.

\section{Construction of geometric objects in the \newline
$\protect\sigma $-immanent form}

In the T-geometry the geometric object $\mathcal{O}$ is described by means
of the skeleton-envelope method \cite{R01}. It means that any geometric
object $\mathcal{O}$ is considered to be a set of intersections and joins of
elementary geometric objects (EGO).

The finite set $\mathcal{P}^{n}\equiv \left\{ P_{0},P_{1},...,P_{n}\right\}
\subset \Omega $ of parameters of the envelope function $f_{\mathcal{P}^{n}}$
is the skeleton of elementary geometric object (EGO) $\mathcal{E}\subset
\Omega $. The set $\mathcal{E}\subset \Omega $ of points forming EGO is
called the envelope of its skeleton $\mathcal{P}^{n}$. In the continuous
generalized geometry the envelope $\mathcal{E}$ is usually a continual set
of points. The envelope function $f_{\mathcal{P}^{n}}$%
\begin{equation}
f_{\mathcal{P}^{n}}:\qquad \Omega \rightarrow \mathbb{R},  \label{h2.1}
\end{equation}%
determining EGO is a function of the running point $R\in \Omega $ and of
parameters $\mathcal{P}^{n}\subset \Omega $. The envelope function $f_{%
\mathcal{P}^{n}}$ is supposed to be an algebraic function of $s$ arguments $%
w=\left\{ w_{1},w_{2},...w_{s}\right\} $, $s=(n+2)(n+1)/2$. Each of
arguments $w_{k}=\sigma \left( Q_{k},L_{k}\right) $ is a $\sigma $-function
of two arguments $Q_{k},L_{k}\in \left\{ R,\mathcal{P}^{n}\right\} $, either
belonging to skeleton $\mathcal{P}^{n}$, or coinciding with the running
point $R$. Thus, any elementary geometric object $\mathcal{E}$ is determined
by its skeleton and its envelope function as the set of zeros of the
envelope function 
\begin{equation}
\mathcal{E}=\left\{ R|f_{\mathcal{P}^{n}}\left( R\right) =0\right\}
\label{h2.2}
\end{equation}

For instance, the cylinder $\mathcal{C}(P_{0},P_{1},Q)$ with the points $%
P_{0},P_{1}$ on the cylinder axis and the point $Q$ on its surface is
determined by the relation 
\begin{eqnarray}
\mathcal{C}(P_{0},P_{1},Q) &=&\left\{ R|f_{P_{0}P_{1}Q}\left( R\right)
=0\right\} ,  \label{g3.1} \\
f_{P_{0}P_{1}Q}\left( R\right) &=&F_{2}\left( P_{0},P_{1},Q\right)
-F_{2}\left( P_{0},P_{1},R\right)  \nonumber
\end{eqnarray}
\begin{equation}
F_{2}\left( P_{0},P_{1},Q\right) =\left\vert 
\begin{array}{cc}
\left( \mathbf{P}_{0}\mathbf{P}_{1}.\mathbf{P}_{0}\mathbf{P}_{1}\right) & 
\left( \mathbf{P}_{0}\mathbf{P}_{1}.\mathbf{P}_{0}\mathbf{Q}\right) \\ 
\left( \mathbf{P}_{0}\mathbf{Q}.\mathbf{P}_{0}\mathbf{P}_{1}\right) & \left( 
\mathbf{P}_{0}\mathbf{Q}.\mathbf{P}_{0}\mathbf{Q}\right)%
\end{array}
\right\vert  \label{g3.2}
\end{equation}
Here $\sqrt{F_{2}\left( P_{0},P_{1},Q\right) }$ is the area of the
parallelogram, constructed on the vectors $\mathbf{P}_{0}\mathbf{P}_{1}$ and 
$\mathbf{P}_{0}\mathbf{Q}$\textbf{\ }and $\frac{1}{2}\sqrt{F_{2}\left(
P_{0},P_{1},Q\right) }\ $ is the area of triangle with vertices at the
points $P_{0},P_{1},Q$. The equality $F_{2}\left( P_{0},P_{1},Q\right)
=F_{2}\left( P_{0},P_{1},R\right) $ means that the distance between the
point $Q$ and the axis, determined by the vector $\mathbf{P}_{0}\mathbf{P}%
_{1}$ is equal to the distance between $R$ and the axis.

The elementary geometrical object $\mathcal{E}$ is determined in all
physical geometries at once. In particular, it is determined in the proper
Euclidean geometry, where we can obtain its meaning. We interpret the
elementary geometrical object $\mathcal{E}$, using our knowledge of the
proper Euclidean geometry. Thus, the proper Euclidean geometry is used as a
sample geometry for interpretation of any physical geometry. In particular,
the cylinder (\ref{g3.1}) is determined uniquely in any T-geometry with any
world function $\sigma .$

In the Euclidean geometry the points $P_{0}$ and $P_{1}$ determine the
cylinder axis. The shape of a cylinder depends on its axis and radius, but
not on the disposition of points $P_{0},P_{1}$ on the cylinder axis. As a
result in the Euclidean geometry the cylinders $\mathcal{C}(P_{0},P_{1},Q)$
and $\mathcal{C}(P_{0},P_{2},Q)$ coincide, provided vectors $\mathbf{P}_{0}%
\mathbf{P}_{1}$ and $\mathbf{P}_{0}\mathbf{P}_{2}$ are collinear. In the
general case of T-geometry the cylinders $\mathcal{C}(P_{0},P_{1},Q)$ and $%
\mathcal{C}(P_{0},P_{2},Q)$ do not coincide, in general, even if vectors $%
\mathbf{P}_{0}\mathbf{P}_{1}$ and $\mathbf{P}_{0}\mathbf{P}_{2}$ are
collinear. Thus, in general, a deformation of the Euclidean geometry splits
Euclidean geometrical objects.

We do not try to repeat subscriptions of Euclid at construction of the
geometry. We take the geometrical objects and relations between them,
prepared in the framework of the Euclidean geometry and describe them in
terms of the world function. Thereafter we deform them, replacing the
Euclidean world function $\sigma _{\mathrm{E}}$ by the world function $%
\sigma $ of the geometry in question. In practice the construction of the
elementary geometric object is reduced to the representation of the
corresponding Euclidean geometrical object in the $\sigma $-immanent form,
i.e. in terms of the Euclidean world function. The last problem is the
problem of the proper Euclidean geometry. The problem of representation of
the geometrical object (or relation between objects) in the $\sigma $%
-immanent form is a real problem of the T-geometry construction.

Application of the deformation principle is restricted by two constraints.

1. The deformation principle is to be applied separately from other methods
of the geometry construction. In particular, one may not use topological
structures in construction of a physical geometry, because for effective
application of the deformation principle the obtained physical geometry must
be determined only by the world function (metric).

2. Describing Euclidean geometric objects $\mathcal{O}\left( \sigma _{%
\mathrm{E}}\right) $ and Euclidean relation $\mathcal{R}\left( \sigma _{%
\mathrm{E}}\right) $ in terms of $\sigma _{\mathrm{E}}$, we are not to use
special properties of Euclidean world function $\sigma _{\mathrm{E}}$. In
particular, definitions of $\mathcal{O}\left( \sigma _{\mathrm{E}}\right) $
and $\mathcal{R}\left( \sigma _{\mathrm{E}}\right) $ are to have similar
form in Euclidean geometries of different dimensions. They must not depend
on the dimension of the Euclidean space.

The T-geometry construction is not to use coordinates and other methods of
description, because the application of the means of description imposes
constraints on the constructed geometry. Any means of description is a
structure $St$ given on the basic Euclidean geometry with the world function 
$\sigma _{\mathrm{E}}$. Replacement $\sigma _{\mathrm{E}}\rightarrow \sigma $
is sufficient for construction of unique generalized geometry $\mathcal{G}%
_{\sigma }$. If we use an additional structure $St$ for the T-geometry
construction, we obtain, in general, other geometry $\mathcal{G}_{St}$,
which coincides with $\mathcal{G}_{\sigma }$ not for all $\sigma $, but only
for some of world functions $\sigma $. Thus, a use of additional means of
description restricts the list of possible generalized geometries. For
instance, if we use the coordinate description at construction of the
generalized geometry, the obtained geometry appears to be continuous,
because description by means of the coordinates is effective only for
continuous geometries, where the number of coordinates coincides with the
geometry dimension.

As far as the $\sigma $-immanent description of the proper Euclidean
geometry is possible, it is possible for any T-geometry, because any
geometrical object $\mathcal{O}$ and any relation $\mathcal{R}$ in the
physical geometry $\mathcal{G}$ is obtained from the corresponding
geometrical object $\mathcal{O}_{\mathrm{E}}$ and from the corresponding
relation $\mathcal{R}_{\mathrm{E}}$ in the proper Euclidean geometry $%
\mathcal{G}_{\mathrm{E}}$ by means of the replacement $\sigma _{\mathrm{E}%
}\rightarrow \sigma $ in description of $\mathcal{O}_{\mathrm{E}}$ and $%
\mathcal{R}_{\mathrm{E}}$. For such a replacement be possible, the
description of $\mathcal{O}_{\mathrm{E}}$ and $\mathcal{R}_{\mathrm{E}}$ is
not to refer to special properties of $\sigma _{\mathrm{E}}$, described by
conditions II -- V. A formal indicator of the conditions II -- V application
is a reference to the dimension $n$, because any of conditions II -- V
contains a reference to the dimension $n$ of the proper Euclidean space.

Let us suppose that some geometrical object $\mathcal{O}_{\mathrm{E}%
_{n}}\left( \sigma _{\mathrm{E}_{n}},n\right) $ is defined in the $n$%
-dimensional Euclidean space, and this definition refers explicitly to the
dimension of the Euclidean space $n$. Let us deform the $n$-dimensional
Euclidean space $E_{n}$ in the $m$-dimensional Euclidean space $E_{m}$. Then
we must make the change 
\begin{equation}
\mathcal{O}_{\mathrm{E}_{n}}\left( \sigma _{\mathrm{E}_{n}},n\right)
\rightarrow \mathcal{O}_{\mathrm{E}_{n}}\left( \sigma _{\mathrm{E}%
_{m}},n\right)  \label{g3.3}
\end{equation}
On the other hand, we may define the same geometrical object directly in the 
$m$-dimensional Euclidean space $E_{m}$ in the form $\mathcal{O}_{\mathrm{E}%
_{m}}\left( \sigma _{\mathrm{E}_{m}},m\right) $. Equating this expression to
(\ref{g3.3}), we obtain 
\begin{equation}
\mathcal{O}_{\mathrm{E}_{n}}\left( \sigma _{\mathrm{E}_{m}},n\right) =%
\mathcal{O}_{\mathrm{E}_{m}}\left( \sigma _{\mathrm{E}_{m}},m\right) ,\qquad
\forall m,n\in \mathbb{N}  \label{g3.4}
\end{equation}

It means that the definition of the geometrical object $\mathcal{O}$ is to
be independent on the dimension of the Euclidean space.

If nevertheless we use one of special properties II -- V of the Euclidean
space in the $\sigma $-immanent description of a geometrical object $%
\mathcal{O}$, or relation $\mathcal{R}$ , we refer to the dimension $n$ and,
ultimately, to the coordinate system, which is only a means of description.

Let us show this in the example of the determination of the straight in the
Euclidean space. The straight $\mathcal{T}_{P_{0}Q}$ in the proper Euclidean
space is defined by two its points $P_{0}$ and $Q$ $\;\left( P_{0}\neq
Q\right) $ as the set of points $R$ 
\begin{equation}
\mathcal{T}_{P_{0}Q}=\left\{ R\;|\;\mathbf{P}_{0}\mathbf{Q}||\mathbf{P}_{0}%
\mathbf{R}\right\}  \label{b15}
\end{equation}%
where condition $\mathbf{P}_{0}\mathbf{Q}||\mathbf{P}_{0}\mathbf{R}$ means
that vectors $\mathbf{P}_{0}\mathbf{Q}$ and $\mathbf{P}_{0}\mathbf{R}$ are
collinear, i.e. the scalar product $\left( \mathbf{P}_{0}\mathbf{Q}.\mathbf{P%
}_{0}\mathbf{R}\right) $ of these two vectors satisfies the relation 
\begin{equation}
\mathbf{P}_{0}\mathbf{Q}||\mathbf{P}_{0}\mathbf{R:\qquad }\left( \mathbf{P}%
_{0}\mathbf{Q}.\mathbf{P}_{0}\mathbf{R}\right) ^{2}=\left( \mathbf{P}_{0}%
\mathbf{Q}.\mathbf{P}_{0}\mathbf{Q}\right) \left( \mathbf{P}_{0}\mathbf{R}.%
\mathbf{P}_{0}\mathbf{R}\right)  \label{b16}
\end{equation}%
where the scalar product is defined by the relation (\ref{b1.1a}). Thus, the
straight line $\mathcal{T}_{P_{0}Q}$ is defined $\sigma $-immanently, i.e.
in terms of the world function $\sigma $. We shall use two different names
(straight and tube) for the geometric object $\mathcal{T}_{P_{0}Q}$. We
shall use the term \textquotedblright straight\textquotedblright , when we
want to stress that $\mathcal{T}_{P_{0}Q}$ is a result of deformation of the
Euclidean straight. We shall use the term \textquotedblright
tube\textquotedblright , when we want to stress that $\mathcal{T}_{P_{0}Q}$
may be a many-dimensional surface.

In the Euclidean geometry one can use another definition of collinearity.
Vectors $\mathbf{P}_{0}\mathbf{Q}$ and $\mathbf{P}_{0}\mathbf{R}$ are
collinear, if components of vectors $\mathbf{P}_{0}\mathbf{Q}$ and $\mathbf{P%
}_{0}\mathbf{R}$ are proportional in some rectilinear coordinate system. For
instance, in the $n$-dimensional Euclidean space one can introduce
rectilinear coordinate system, choosing $n+1$ points $\mathcal{P}%
^{n}=\left\{ P_{0},P_{1},...P_{n}\right\} $ and forming $n$ basic vectors $%
\mathbf{P}_{0}\mathbf{P}_{i}$, $i=1,2,...n$. Then the collinearity condition
can be written in the form of $n$ equations 
\begin{equation}
\mathbf{P}_{0}\mathbf{Q}||\mathbf{P}_{0}\mathbf{R:\qquad }\left( \mathbf{P}%
_{0}\mathbf{P}_{i}.\mathbf{P}_{0}\mathbf{Q}\right) =a\left( \mathbf{P}_{0}%
\mathbf{P}_{i}.\mathbf{P}_{0}\mathbf{R}\right) ,\qquad i=1,2,...n,\qquad
a\in \mathbb{R}\backslash \left\{ 0\right\}  \label{b17}
\end{equation}%
where $a\neq 0$ is some real constant. Relations (\ref{b17}) are relations
for covariant components of vectors $\mathbf{P}_{0}\mathbf{Q}$ and $\mathbf{P%
}_{0}\mathbf{R}$ in the considered coordinate system with basic vectors $%
\mathbf{P}_{0}\mathbf{P}_{i}$, $i=1,2,...n$. The definition of collinearity (%
\ref{b17}) depends on the dimension $n$ of the Euclidean space. Let points $%
\mathcal{P}^{n}$ be chosen in such a way, that $\left( \mathbf{P}_{0}\mathbf{%
P}_{1}.\mathbf{P}_{0}\mathbf{Q}\right) \neq 0$. Then eliminating the
parameter $a$ from relations (\ref{b17}), we obtain $n-1$ independent
relations, and the geometrical object 
\begin{eqnarray}
\mathcal{T}_{Q\mathcal{P}^{n}} &=&\left\{ R\;|\;\mathbf{P}_{0}\mathbf{Q}||%
\mathbf{P}_{0}\mathbf{R}\right\} =\bigcap\limits_{i=2}^{i=n}\mathcal{S}_{i},
\label{c2.1} \\
\mathcal{S}_{i} &=&\left\{ R\left\vert \frac{\left( \mathbf{P}_{0}\mathbf{P}%
_{i}.\mathbf{P}_{0}\mathbf{Q}\right) }{\left( \mathbf{P}_{0}\mathbf{P}_{1}.%
\mathbf{P}_{0}\mathbf{Q}\right) }=\frac{\left( \mathbf{P}_{0}\mathbf{P}_{i}.%
\mathbf{P}_{0}\mathbf{R}\right) }{\left( \mathbf{P}_{0}\mathbf{P}_{1}.%
\mathbf{P}_{0}\mathbf{R}\right) }\right. \right\} ,\qquad i=2,3,...n
\label{c2.2}
\end{eqnarray}%
defined according to (\ref{b17}), depends on $n+2$ points $Q,\mathcal{P}^{n}$%
. This geometrical object $\mathcal{T}_{Q\mathcal{P}^{n}}$ is defined $%
\sigma $-immanently. It is a complex, consisting of the straight line and of
the coordinate system, represented by $n+1$ points $\mathcal{P}^{n}=\left\{
P_{0},P_{1},...P_{n}\right\} $. In the Euclidean space the dependence on the
choice of the coordinate system and on $n$ points $\left\{
P_{1},...P_{n}\right\} $, determining this system, is fictitious. The
geometrical object $\mathcal{T}_{Q\mathcal{P}^{n}}$ depends essentially only
on two points $P_{0},Q$ and coincides with the straight line $\mathcal{T}%
_{P_{0}Q}$ in the Euclidean space. But at deformations of the Euclidean
space the geometrical objects $\mathcal{T}_{Q\mathcal{P}^{n}}$ and $\mathcal{%
T}_{P_{0}Q}$ are deformed differently. The points $P_{1},P_{2},...P_{n}$
cease to be fictitious in definition of $\mathcal{T}_{Q\mathcal{P}^{n}}$,
and geometrical objects $\mathcal{T}_{Q\mathcal{P}^{n}}$ and $\mathcal{T}%
_{P_{0}Q}$ become to be different geometric objects, in general. But being
different, in general, they may coincide in some special cases.

What of the two geometrical objects in the deformed geometry $\mathcal{G}$
should be interpreted as a straight line, passing through the points $P_{0}$
and $Q$ in the geometry $\mathcal{G}$? Of course, it is $\mathcal{T}%
_{P_{0}Q} $, because its definition does not contain a reference to a
coordinate system, whereas definition of $\mathcal{T}_{Q\mathcal{P}^{n}}$
depends on the choice of the coordinate system, represented by points $%
\mathcal{P}^{n}$. In general, definitions of geometric objects and relations
between them are not to refer to the means of description. Otherwise, the
points determining the coordinate system are to be included in definition of
the geometrical object.

But in the given case the geometrical object $\mathcal{T}_{P_{0}Q}$ is a $%
(n-1)$-dimensional surface, in general, whereas $\mathcal{T}_{Q\mathcal{P}%
^{n}}$ is an intersection of $(n-1)\;\;$ $(n-1)$-dimensional surfaces, i.e. $%
\mathcal{T}_{Q\mathcal{P}^{n}}$ is a one-dimensional curve, in general. The
one-dimensional curve $\mathcal{T}_{Q\mathcal{P}^{n}}$ corresponds better to
our ideas on the straight line, than the $(n-1)$-dimensional surface $%
\mathcal{T}_{P_{0}Q}$. Nevertheless, in physical geometry $\mathcal{G}$ it
is $\mathcal{T}_{P_{0}Q}$, that is an analog of the Euclidean straight line.

It is very difficult to overcome our conventional idea that the Euclidean
straight line cannot be deformed into many-dimensional surface, and \textit{%
this idea has been prevent for years from construction of T-geometries}.
Practically one uses such physical geometries, where deformation of the
Euclidean space transforms the Euclidean straight lines into one-dimensional
lines. It means that one chooses such geometries, where geometrical objects $%
\mathcal{T}_{P_{0}Q}$ and $\mathcal{T}_{Q\mathcal{P}^{n}}$ coincide. 
\begin{equation}
\mathcal{T}_{P_{0}Q}=\mathcal{T}_{Q\mathcal{P}^{n}}  \label{b19}
\end{equation}%
Condition (\ref{b19}) of coincidence of the objects $\mathcal{T}_{P_{0}Q}$
and $\mathcal{T}_{Q\mathcal{P}^{n}}$, imposed on the T-geometry, restricts
the list of possible T-geometries.

In general, the condition (\ref{b19}) cannot be fulfilled, because lhs does
not depend on points $\left\{ P_{1},P_{2},...P_{n}\right\} $, whereas rhs of
(\ref{b19}) depends, in general. The tube $\mathcal{T}_{Q\mathcal{P}^{n}}$
does not depend on the points $\left\{ P_{1},P_{2},...P_{n}\right\} $,
provided the distance $\sqrt{2\sigma \left( P_{i},P_{k}\right) }$ between
any two points $P_{i},P_{k}\in \mathcal{P}^{n}$ is infinitesimal. In the
Riemannian geometry the constraint (\ref{b19}) is fulfilled at the
additional restriction.%
\begin{equation}
\sqrt{2\sigma \left( P_{i},P_{k}\right) }=\text{infinitesimal},\qquad
i,k=1,2,...n  \label{b20}
\end{equation}

\section{Interplay between metric geometry and \newline
T-geometry}

Let us consider the metric geometry, given on the set $\Omega $ of points.
The metric space $M=\left\{ \rho ,\Omega \right\} $ is given by the metric
(distance) $\rho $. 
\begin{eqnarray}
\rho &:&\quad \Omega \times \Omega \rightarrow \lbrack 0,\infty )\subset 
\mathbb{R}  \label{c2.3} \\
\rho (P,P) &=&0,\qquad \rho (P,Q)=\rho (Q,P),\qquad \forall P,Q\in \Omega
\label{c2.4} \\
\rho (P,Q) &\geq &0,\qquad \rho (P,Q)=0,\quad \text{iff }P=Q,\qquad \forall
P,Q\in \Omega  \label{c2.5} \\
0 &\leq &\rho (P,R)+\rho (R,Q)-\rho (P,Q),\qquad \forall P,Q,R\in \Omega
\label{c2.6}
\end{eqnarray}%
At first sight the metric space is a special case of the $\sigma $-space (%
\ref{h1.1}), and the metric geometry is a special case of the T-geometry
with additional constraints (\ref{c2.5}), (\ref{c2.6}) imposed on the world
function $\sigma =\frac{1}{2}\rho ^{2}$. However it is not so, because the
metric geometry is not equipped by the deformation principle. The fact, that
the $\sigma $-immanence of the Euclidean geometry, as well as the complex of
conditions (\ref{a1.4}) - (\ref{b14}), was not known until 1990, although
any of relations (\ref{a1.4}) - (\ref{b14}) was well known. Additional (with
respect to the $\sigma $-space) constraints (\ref{c2.5}), (\ref{c2.6}) are
imposed to provide one-dimensionality of the straight lines. In the metric
geometry the shortest (straight) line can be constructed only in the case,
when it is one-dimensional.

Let us consider the set $\mathcal{EL}\left( P,Q,a\right) $ of points $R$ 
\begin{equation}
\mathcal{EL}\left( P,Q,2a\right) =\left\{ R|f_{P,Q,2a}\left( R\right)
=0\right\} ,\qquad f_{P,Q,2a}\left( R\right) =\rho (P,R)+\rho (R,Q)-2a
\label{c2.8}
\end{equation}
If the metric space coincides with the proper Euclidean space, this set of
points is an ellipsoid with focuses at the points $P,Q$ and the large
semiaxis $a$. The relations $f_{P,Q,2a}\left( R\right) >0$, $%
f_{P,Q,2a}\left( R\right) =0$, $f_{P,Q,2a}\left( R\right) <0$ determine
respectively external points, boundary points and internal points of the
ellipsoid. If $\rho \left( P,Q\right) =2a$, we obtain the degenerate
ellipsoid, which coincides with the segment $\mathcal{T}_{\left[ PQ\right] }$
of the straight line, passing through the points $P$, $Q$.\ In the proper
Euclidean geometry, the degenerate ellipsoid is one-dimensional segment of
the straight line, but it is not evident that it is one-dimensional in the
case of arbitrary metric geometry. For such a degenerate ellipsoid be
one-dimensional in the arbitrary metric space, it is necessary that any
degenerate ellipsoid $\mathcal{EL}\left( P,Q,\rho \left( P,Q\right) \right) $
have no internal points. This constraint is written in the form 
\begin{equation}
f_{P,Q,\rho \left( P,Q\right) }\left( R\right) =\rho (P,R)+\rho (R,Q)-\rho
(P,Q)\geq 0  \label{c2.9}
\end{equation}

Comparing relation (\ref{c2.9}) with (\ref{c2.6}), we see that the
constraint (\ref{c2.6}) is introduced to provide the straight (shortest)
line one-dimensionality (absence of internal points in the geometrical
object determined by two points).

As far as the metric geometry does not use the deformation principle, it is
a poor geometry, because in the framework of this geometry one cannot
construct the scalar product of two vectors, define linear independence of
vectors and construct such geometrical objects as planes. All these objects
as well as other are constructed on the basis of the deformation of the
proper Euclidean geometry.

Generalizing the metric geometry, Menger \cite{M28} and Blumenthal \cite{B53}
removed the triangle axiom (\ref{c2.6}). They tried to construct the
distance geometry, which would be a more general geometry, than the metric
one. As far as they did not use the deformation principle, they could not
determine the shortest (straight) line without a reference to the
topological concept of the curve $\mathcal{L}$, defined as a continuous
mapping (\ref{g1.1}), which cannot be expressed only via the distance. As a
result the distance geometry appeared to be not a pure metric geometry (i.e.
the geometry determined only by the distance).

Note that the Riemannian geometry uses the deformation principle in the
coordinate form. The distance geometry cannot use it in such a form, because
the metric and distance geometries are formulated in the coordinateless
form. It is to use the deformation principle in the coordinateless form. But
application of the deformation principle in the coordinateless form needs a
use of the Euclidean geometry $\sigma $-immanence. K. Menger went to the
concept of the $\sigma $-immanence, but he stopped in one step before the $%
\sigma $-immanence. Look at the K. Menger's theorem \cite{M28}, written in
our designations

\begin{theorem}
The $\sigma $-space $V=\left\{ \sigma ,\Omega \right\} $ is
isometrically embeddable in $n$-dimensional proper Euclidean space
$E_{n}$, if and only if any set of $n+3$ points of $\Omega $ is
isometrically embeddable in $E_{n}$.
\end{theorem}

The theorem on the $\sigma $-immanence of the Euclidean geometry is obtained
from the Menger's theorem, if instead of the condition "any set of $n+3$
points of $\Omega $ is isometrically embeddable in $E_{n}"$ one writes the
condition (\ref{a1.5a}), which also contains $n+3$ points: $P,Q,\mathcal{P}%
^{n}$ and describes the fact that $\left\{ P,Q,\mathcal{P}^{n}\right\}
\subset E_{n}$. In this case the theorem condition contains only a reference
to the properties of the world function of the Euclidean space, but not to
the Euclideaness of the space. (continuity of the $\sigma $-space $V$ is
neglected in such a formulation.)

\section{Conditions of the deformation principle \newline
application}

Riemannian geometries satisfy the condition (\ref{b19}). The Riemannian
geometry is a kind of inhomogeneous physical geometry, and, hence, it uses
the deformation principle. Constructing the Riemannian geometry, the
infinitesimal Euclidean distance is deformed into the Riemannian distance.
The deformation is chosen in such a way that any Euclidean straight line $%
\mathcal{T}_{\mathrm{E}P_{0}Q}$, passing through the point $P_{0}$,
collinear to the vector $\mathbf{P}_{0}\mathbf{Q}$, is transformed into the
geodesic $\mathcal{T}_{P_{0}Q}$, passing through the point $P_{0}$,
collinear to the vector $\mathbf{P}_{0}\mathbf{Q}$ in the Riemannian space.

Note that in T-geometries, satisfying the condition (\ref{b19}) for all
points $Q,\mathcal{P}^{n}$, the straight line 
\begin{equation}
\mathcal{T}_{Q_{0};P_{0}Q}=\left\{ R\;|\;\mathbf{P}_{0}\mathbf{Q}||\mathbf{Q}%
_{0}\mathbf{R}\right\}  \label{b3.0}
\end{equation}%
passing through the point $Q_{0}$ collinear to the vector $\mathbf{P}_{0}%
\mathbf{Q}$, is not a one-dimensional line, in general. If the Riemannian
geometries be T-geometries, they would contain non-one-dimensional geodesics
(straight lines). But the Riemannian geometries are not T-geometries,
because at their construction one uses not only the deformation principle,
but some other methods, containing a reference to the means of description.
In particular, in the Riemannian geometries the absolute parallelism is
absent, and one cannot define a straight line (\ref{b3.0}), because the
collinearity relation $\mathbf{P}_{0}\mathbf{Q}||\mathbf{Q}_{0}\mathbf{R}$
is not defined, if points $P_{0}$ and $Q_{0}$ do not coincide. On one hand,
a lack of absolute parallelism allows one to go around the problem of
non-one-dimensional straight lines. On the other hand, it makes the
Riemannian geometries to be inconsistent, because they cease to be
T-geometries, which are consistent by the construction (see for details \cite%
{R02}).

The fact is that the application of \textit{only deformation principle }is
sufficient for construction of a physical geometry. Besides, such a
construction is consistent, because the original Euclidean geometry is
consistent and, deforming it, we do not use any logical reasonings. If we
introduce additional structure (for instance, a topological structure) we
obtain a fortified physical geometry, i.e. a physical geometry with
additional structure on it. The physical geometry, equipped with additional
structure, is a more pithy construction, than the physical geometry simply.
But it is valid only in the case, when we consider the additional structure
as an addition to the physical geometry. If we use an additional structure
in construction of the geometry, we identify the additional structure with
one of structures of the physical geometry. If we demand that the additional
structure be a structure of physical geometry, we restrict an application of
the deformation principle and reduce the list of possible physical
geometries, because coincidence of the additional structure with some
structure of a physical geometry is possible not for all physical
geometries, but only for some of them.

Let, for instance, we use concept of a curve $\mathcal{L}$ (\ref{g1.1}) for
construction of a physical geometry. The concept of curve $\mathcal{L}$,
considered as a continuous mapping, is a topological structure, which cannot
be expressed only via the distance or via the world function. A use of the
mapping (\ref{g1.1}) needs an introduction of topological space and, in
particular, the concept of continuity. If we identify the topological curve (%
\ref{g1.1}) with the \textquotedblright metrical\textquotedblright\ curve,
defined as a broken line 
\begin{equation}
\mathcal{T}_{\mathrm{br}}=\bigcup\limits_{i}\mathcal{T}_{\left[ P_{i}P_{i+1}%
\right] },\qquad \mathcal{T}_{\left[ P_{i}P_{i+1}\right] }=\left\{ R|\sqrt{%
2\sigma \left( P_{i},P_{i+1}\right) }-\sqrt{2\sigma \left( P_{i},R\right) }-%
\sqrt{2\sigma \left( R,P_{i+1}\right) }\right\}  \label{a1.2}
\end{equation}%
consisting of the straight line segments $\mathcal{T}_{\left[ P_{i}P_{i+1}%
\right] }$ between the points $P_{i}$, $P_{i+1}$, we truncate the list of
possible geometries, because such an identification is possible only in some
physical geometries. Identifying (\ref{g1.1}) and (\ref{a1.2}), we eliminate
all discrete physical geometries and those continuous physical geometries,
where the segment $\mathcal{T}_{\left[ P_{i}P_{i+1}\right] }$ of straight
line is a surface, but not a one-dimensional set of points. Thus, additional
structures may lead to (i) a fortified physical geometry, (ii) a restricted
physical geometry and (iii) a restricted fortified physical geometry. The
result depends on the method of the additional structure application.

Note that some constraints (continuity, convexity, lack of absolute
parallelism), imposed on physical geometries are a result of a disagreement
of the means of description, which are used at the geometry construction. In
the T-geometry, which uses only the deformation principle, there is no such
restrictions. Besides, the T-geometry has some new property of a physical
geometry, which is not accepted by conventional versions of physical
geometry. This property, called the geometry nondegeneracy, follows directly
from the application of arbitrary deformations to the proper Euclidean
geometry.

\begin{definition}
The geometry is degenerate at the point $P_{0}$ in the direction of the
vector $\mathbf{Q}_{0}\mathbf{Q}$, $\left\vert \mathbf{Q}_{0}\mathbf{Q}%
\right\vert \neq 0$, if the relations
\begin{equation}
\mathbf{Q}_{0}\mathbf{Q}\uparrow \uparrow \mathbf{P}_{0}\mathbf{R:\qquad }%
\left( \mathbf{Q}_{0}\mathbf{Q}.\mathbf{P}_{0}\mathbf{R}\right) =\sqrt{%
\left\vert \mathbf{Q}_{0}\mathbf{Q}\right\vert \cdot \left\vert \mathbf{P}%
_{0}\mathbf{R}\right\vert },\qquad \left\vert \mathbf{P}_{0}\mathbf{R}%
\right\vert =a\neq 0  \label{b3.1}
\end{equation}%
considered as equations for determination of the point $R$, have not more,
than one solution for any $a\neq 0$. Otherwise, the geometry is
nondegenerate at the point $P_{0}$ in the direction of the vector $\mathbf{Q}%
_{0}\mathbf{Q}$.
\end{definition}

Note that the first equation (\ref{b3.1}) is the condition of the
parallelism of vectors $\mathbf{Q}_{0}\mathbf{Q}$ and $\mathbf{P}_{0}\mathbf{%
R}$.

The proper Euclidean geometry is degenerate, i.e. it is degenerate at all
points in directions of all vectors. Considering the Minkowski geometry, one
should distinguish between the Minkowski T-geometry and Minkowski geometry.
The two geometries are described by the same world function and differ in
the definition of the parallelism. In the Minkowski T-geometry the
parallelism of two vectors $\mathbf{\mathbf{Q}_{0}\mathbf{Q}}$ and $\mathbf{%
\mathbf{P}_{0}\mathbf{R}}$ is defined by the first equation (\ref{b3.1}).
This definition is based on the deformation principle. In the $n$%
-dimensional Minkowski geometry ($n$-dimensional pseudo-Euclidean geometry
of index $1$) the parallelism is defined by the relation of the type of (\ref%
{b17}) 
\begin{equation}
\mathbf{Q}_{0}\mathbf{Q}\uparrow \uparrow \mathbf{P}_{0}\mathbf{R:\qquad }%
\left( \mathbf{P}_{0}\mathbf{P}_{i}.\mathbf{Q}_{0}\mathbf{Q}\right) =a\left( 
\mathbf{P}_{0}\mathbf{P}_{i}.\mathbf{P}_{0}\mathbf{R}\right) ,\qquad
i=1,2,...n,\qquad a>0  \label{b3.1a}
\end{equation}%
where points $\mathcal{P}^{n}=\left\{ P_{0},P_{1},...P_{n}\right\} $
determine a rectilinear coordinate system with basic vectors $\mathbf{P}_{0}%
\mathbf{P}_{i}$, $i=1,2,..n$ in the $n$-dimensional Minkowski geometry .
Dependence of the definition (\ref{b3.1a}) on the points $\left(
P_{0},P_{1},...P_{n}\right) $ is fictitious, but dependence on the number $%
n+1$ of points $\mathcal{P}^{n}$ is essential. Thus, definition (\ref{b3.1a}%
) depends on the method of the geometry description.

The Minkowski T-geometry is degenerate at all points in direction of all
timelike vectors, and it is nondegenerate at all points in direction of all
spacelike vectors. The Minkowski geometry is degenerate at all points in
direction of all vectors. Conventionally one uses the Minkowski geometry,
ignoring the nondegeneracy in spacelike directions.

Considering the proper Riemannian geometry, one should distinguish between
the Riemannian T-geometry and the Riemannian geometry. The two geometries
are described by the same world function. They differ in the definition of
the parallelism. In the Riemannian T-geometry the parallelism of two vectors 
$\mathbf{\mathbf{Q}_{0}\mathbf{Q}}$ and $\mathbf{\mathbf{P}_{0}\mathbf{R}}$
is defined by (\ref{b3.1}). In the Riemannian geometry the parallelism of
two vectors $\mathbf{\mathbf{Q}_{0}\mathbf{Q}}$ and $\mathbf{\mathbf{P}_{0}%
\mathbf{R}}$ is defined only in the case, when the points $P_{0}$ and $Q_{0}$
coincide. Parallelism of remote vectors $\mathbf{\mathbf{Q}_{0}\mathbf{Q}}$
and $\mathbf{\mathbf{P}_{0}\mathbf{R}}$ is not defined, in general. This
fact is known as absence of absolute parallelism.

The proper Riemannian T-geometry is locally degenerate, i.e. it is
degenerate at all points $P_{0}$ in direction of all vectors $\mathbf{P}_{0}%
\mathbf{Q}$ with the origin at the point $P_{0}$. In the general case, when $%
P_{0}\neq Q_{0}$, the proper Riemannian T-geometry is nondegenerate, in
general. But it is degenerate locally as well as the proper Riemannian
geometry. The proper Riemannian geometry is degenerate, because it is
degenerate locally, whereas the nonlocal degeneracy is not defined in the
Riemannian geometry, because of the lack of absolute parallelism.
Conventionally one uses the Riemannian geometry (not Riemannian T-geometry)
and ignores the property of the nondegeneracy completely.

From the viewpoint of the conventional approach to the physical geometry the
nondegeneracy is an undesirable property of a physical geometry, although
from the logical viewpoint and from viewpoint of the deformation principle
the nondegeneracy is \textit{an inherent property of a physical geometry}.
The nonlocal nondegeneracy is ejected from the proper Riemannian geometry by
denial of existence of the remote vector parallelism. Nondegeneracy in the
spacelike directions is ejected from the Minkowski geometry by means of the
redefinition of the two vectors parallelism. But the nondegeneracy is an
important property of the real space-time geometry. To appreciate this, let
us consider an example.

\section{Simple example of nondegenerate space-time geometry}

Let the space-time geometry $\mathcal{G}_{\mathrm{d}}$ be described by the
T-geometry, given on 4-dimensional manifold $\mathcal{M}_{1+3}$. The world
function $\sigma _{\mathrm{d}}$ is described by the relation 
\begin{equation}
\sigma _{\mathrm{d}}=\sigma _{\mathrm{M}}+D\left( \sigma _{\mathrm{M}%
}\right) =\left\{ 
\begin{array}{ll}
\sigma _{\mathrm{M}}+d & \text{if\ }\sigma _{0}<\sigma _{\mathrm{M}} \\ 
\left( 1+\frac{d}{\sigma _{0}}\right) \sigma _{\mathrm{M}} & \text{if\ }%
0\leq \sigma _{\mathrm{M}}\leq \sigma _{0} \\ 
\sigma _{\mathrm{M}} & \text{if\ }\sigma _{\mathrm{M}}<0%
\end{array}%
\right.  \label{b3.3}
\end{equation}%
where $d\geq 0$ and $\sigma _{0}>0$ are some constants. The quantity $\sigma
_{\mathrm{M}}$ is the world function in the Minkowski space-time geometry $%
\mathcal{G}_{\mathrm{M}}$. In the orthogonal rectilinear (inertial)
coordinate system $x=\left\{ t,\mathbf{x}\right\} $ the world function $%
\sigma _{\mathrm{M}}$ has the form 
\begin{equation}
\sigma _{\mathrm{M}}\left( x,x^{\prime }\right) =\frac{1}{2}\left(
c^{2}\left( t-t^{\prime }\right) ^{2}-\left( \mathbf{x}-\mathbf{x}^{\prime
}\right) ^{2}\right)  \label{b3.4}
\end{equation}%
where $c$ is the speed of the light.

Let us compare the broken line (\ref{a1.2}) in Minkowski space-time geometry 
$\mathcal{G}_{\mathrm{M}}$ and in the distorted geometry $\mathcal{G}_{%
\mathrm{d}}$. We suppose that $\mathcal{T}_{\mathrm{br}}$ is timelike broken
line, and all links $\mathcal{T}_{\left[ P_{i}P_{i+1}\right] }$ of $\mathcal{%
T}_{\mathrm{br}}$ are timelike and have the same length 
\begin{equation}
\left\vert \mathbf{P}_{i}\mathbf{P}_{i+1}\right\vert _{\mathrm{d}}=\sqrt{%
2\sigma _{\mathrm{d}}\left( P_{i},P_{i+1}\right) }=\mu _{\mathrm{d}%
}>0,\qquad i=0,\pm 1,\pm 2,...  \label{b3.5}
\end{equation}%
\begin{equation}
\left\vert \mathbf{P}_{i}\mathbf{P}_{i+1}\right\vert _{\mathrm{M}}=\sqrt{%
2\sigma _{\mathrm{M}}\left( P_{i},P_{i+1}\right) }=\mu _{\mathrm{M}%
}>0,\qquad i=0,\pm 1,\pm 2,...  \label{b3.5a}
\end{equation}%
where indices \textquotedblright d\textquotedblright\ and \textquotedblright
M\textquotedblright\ mean that the quantity is calculated by means of $%
\sigma _{\mathrm{d}}$ and $\sigma _{\mathrm{M}}$ respectively. Vector $%
\mathbf{P}_{i}\mathbf{P}_{i+1}$ is regarded as the momentum of the particle
at the segment $\mathcal{T}_{\left[ P_{i}P_{i+1}\right] }$, and the quantity 
$\left\vert \mathbf{P}_{i}\mathbf{P}_{i+1}\right\vert =\mu $ is interpreted
as its (geometric) mass. It follows from definition (\ref{b1.1a}) and
relation (\ref{b3.3}), that for timelike vectors $\mathbf{P}_{i}\mathbf{P}%
_{i+1}$ with $\mu >\sqrt{2\sigma _{0}}$%
\begin{equation}
\left\vert \mathbf{P}_{i}\mathbf{P}_{i+1}\right\vert _{\mathrm{d}}^{2}=\mu _{%
\mathrm{d}}^{2}=\mu _{\mathrm{M}}^{2}+2d,\qquad \mu _{\mathrm{M}%
}^{2}>2\sigma _{0}  \label{b3.6}
\end{equation}%
\begin{equation}
\left( \mathbf{P}_{i-1}\mathbf{P}_{i}.\mathbf{P}_{i}\mathbf{P}_{i+1}\right)
_{\mathrm{d}}=\left( \mathbf{P}_{i-1}\mathbf{P}_{i}.\mathbf{P}_{i}\mathbf{P}%
_{i+1}\right) _{\mathrm{M}}+d  \label{b3.7}
\end{equation}%
Calculation of the shape of the segment $\mathcal{T}_{\left[ P_{0}P_{1}%
\right] }\left( \sigma _{\mathrm{d}}\right) $ in $\mathcal{G}_{\mathrm{d}}$
gives the relation 
\begin{equation}
r^{2}(\tau )=\left\{ 
\begin{array}{ll}
\tau ^{2}\mu _{\mathrm{d}}^{2}\frac{\left( 1-\frac{\tau d}{2\left( \sigma
_{0}+d\right) }\right) ^{2}}{\left( 1-\frac{2d}{\mu _{\mathrm{d}}^{2}}%
\right) }-\frac{\tau ^{2}\mu _{\mathrm{d}}^{2}\sigma _{0}}{\left( \sigma
_{0}+d\right) }, & 0<\tau <\frac{\sqrt{2(\sigma _{0}+d)}}{\mu _{\mathrm{d}}}
\\ 
\frac{3d}{2}+2d\left( \tau -1/2\right) ^{2}\left( 1-\frac{2d}{\mu _{\mathrm{d%
}}^{2}}\right) ^{-1}, & \frac{\sqrt{2(\sigma _{0}+d)}}{\mu _{\mathrm{d}}}%
<\tau <1-\frac{\sqrt{2(\sigma _{0}+d)}}{\mu _{\mathrm{d}}} \\ 
\left( 1-\tau \right) ^{2}\mu _{\mathrm{d}}^{2}\left[ \frac{\left( 1-\frac{%
\left( 1-\tau \right) d}{2\left( \sigma _{0}+d\right) }\right) ^{2}}{\left(
1-\frac{2d}{\mu _{\mathrm{d}}^{2}}\right) }-\frac{\sigma _{0}}{\left( \sigma
_{0}+d\right) }\right] , & 1-\frac{\sqrt{2(\sigma _{0}+d)}}{\mu _{\mathrm{d}}%
}<\tau <1%
\end{array}%
\right. ,  \label{b3.7a}
\end{equation}%
where $r\left( \tau \right) $ is the spatial radius of the segment $\mathcal{%
T}_{\left[ P_{0}P_{1}\right] }\left( \sigma _{\mathrm{d}}\right) $ in the
coordinate system, where points $P_{0}$ and $P_{1}$ have coordinates $%
P_{0}=\left\{ 0,0,0,0\right\} $, $P_{1}=\left\{ \mu _{\mathrm{d}%
},0,0,0\right\} $ and $\tau $ is a parameter along the segment $\mathcal{T}_{%
\left[ P_{0}P_{1}\right] }\left( \sigma _{\mathrm{d}}\right) $, ($\tau
\left( P_{0}\right) =0$, $\tau \left( P_{1}\right) =1$). One can see from (%
\ref{b3.7a}) that the characteristic value of the segment radius is equal to 
$\sqrt{d}$.

Let the broken tube $\mathcal{T}_{\mathrm{br}}$ describe the
\textquotedblright world tube\textquotedblright\ of a free particle. It
means by definition that any link $\mathbf{P}_{i-1}\mathbf{P}_{i}$ is
parallel to the adjacent link $\mathbf{P}_{i}\mathbf{P}_{i+1}$%
\begin{equation}
\mathbf{P}_{i-1}\mathbf{P}_{i}\uparrow \uparrow \mathbf{P}_{i}\mathbf{P}%
_{i+1}:\qquad \left( \mathbf{P}_{i-1}\mathbf{P}_{i}.\mathbf{P}_{i}\mathbf{P}%
_{i+1}\right) -\left\vert \mathbf{P}_{i-1}\mathbf{P}_{i}\right\vert \cdot
\left\vert \mathbf{P}_{i}\mathbf{P}_{i+1}\right\vert =0  \label{b3.8}
\end{equation}%
Definition of parallelism is different in geometries $\mathcal{G}_{\mathrm{M}%
}$ and $\mathcal{G}_{\mathrm{d}}$. As a result links, which are parallel in
the geometry $\mathcal{G}_{\mathrm{M}}$, are not parallel in $\mathcal{G}_{%
\mathrm{d}}$ and vice versa.

Let $\mathcal{T}_{\mathrm{br}}\left( \sigma _{\mathrm{M}}\right) $ describe
the world line of a free particle in the geometry $\mathcal{G}_{\mathrm{M}}$%
. The angle $\vartheta _{\mathrm{M}}$ between the adjacent links in $%
\mathcal{G}_{\mathrm{M}}$ is defined by the relation 
\begin{equation}
\cosh \vartheta _{\mathrm{M}}=\frac{\left( \mathbf{P}_{-1}\mathbf{P}_{0}.%
\mathbf{P}_{0}\mathbf{P}_{1}\right) _{\mathrm{M}}}{\left| \mathbf{P}_{0}%
\mathbf{P}_{1}\right| _{\mathrm{M}}\cdot \left| \mathbf{P}_{-1}\mathbf{P}%
_{0}\right| _{\mathrm{M}}}=1  \label{b3.9}
\end{equation}
The angle $\vartheta _{\mathrm{M}}=0$, and the geometrical object $\mathcal{T%
}_{\mathrm{br}}\left( \sigma _{\mathrm{M}}\right) $ is a timelike straight
line on the manifold $\mathcal{M}_{1+3}$.

Let now $\mathcal{T}_{\mathrm{br}}\left( \sigma _{\mathrm{d}}\right) $
describe the world tube of a free particle in the geometry $\mathcal{G}_{%
\mathrm{d}}$. The angle $\vartheta _{\mathrm{d}}$ between the adjacent links
in $\mathcal{G}_{\mathrm{d}}$ is defined by the relation 
\begin{equation}
\cosh \vartheta _{\mathrm{d}}=\frac{\left( \mathbf{P}_{i-1}\mathbf{P}_{i}.%
\mathbf{P}_{i}\mathbf{P}_{i+1}\right) _{\mathrm{d}}}{\left\vert \mathbf{P}%
_{i}\mathbf{P}_{i+1}\right\vert _{\mathrm{d}}\cdot \left\vert \mathbf{P}%
_{i-1}\mathbf{P}_{i}\right\vert _{\mathrm{d}}}=1  \label{b3.10}
\end{equation}%
The angle $\vartheta _{\mathrm{d}}=0$ also. If we draw the broken tube $%
\mathcal{T}_{\mathrm{br}}\left( \sigma _{\mathrm{d}}\right) $ on the
manifold $\mathcal{M}_{1+3}$, using coordinates of basic points $P_{i}$ and
measure the angle $\vartheta _{\mathrm{dM}}$ between the adjacent links in
the Minkowski geometry $\mathcal{G}_{\mathrm{M}}$, we obtain for the angle $%
\vartheta _{\mathrm{dM}}$ the following relation 
\begin{equation}
\cosh \vartheta _{\mathrm{dM}}=\frac{\left( \mathbf{P}_{i-1}\mathbf{P}_{i}.%
\mathbf{P}_{i}\mathbf{P}_{i+1}\right) _{\mathrm{M}}}{\left\vert \mathbf{P}%
_{i}\mathbf{P}_{i+1}\right\vert _{\mathrm{M}}\cdot \left\vert \mathbf{P}%
_{i-1}\mathbf{P}_{i}\right\vert _{\mathrm{M}}}=\frac{\left( \mathbf{P}_{i-1}%
\mathbf{P}_{i}.\mathbf{P}_{i}\mathbf{P}_{i+1}\right) _{\mathrm{d}}-d}{%
\left\vert \mathbf{P}_{i}\mathbf{P}_{i+1}\right\vert _{\mathrm{d}}^{2}-2d}
\label{b3.11}
\end{equation}%
Substituting the value of $\left( \mathbf{P}_{i-1}\mathbf{P}_{i}.\mathbf{P}%
_{i}\mathbf{P}_{i+1}\right) _{\mathrm{d}}$, taken from (\ref{b3.10}), we
obtain 
\begin{equation}
\cosh \vartheta _{\mathrm{dM}}=\frac{\mu _{\mathrm{d}}^{d}-d}{\mu _{\mathrm{d%
}}^{2}-2d}\approx 1+\frac{d}{\mu _{\mathrm{d}}^{2}},\qquad d\ll \mu _{%
\mathrm{d}}^{2}  \label{b3.12}
\end{equation}%
Hence, $\vartheta _{\mathrm{dM}}\approx \sqrt{2d}/\mu _{\mathrm{d}}$. It
means, that the adjacent link is located on the cone of angle $\sqrt{2d}/\mu
_{\mathrm{d}}$, and the whole line $\mathcal{T}_{\mathrm{br}}\left( \sigma _{%
\mathrm{d}}\right) $ has a random shape, because any link wobbles with the
characteristic angle $\sqrt{2d}/\mu _{\mathrm{d}}$. The wobble angle depends
on the space-time distortion $d$ and on the particle mass $\mu _{\mathrm{d}}$%
. The wobble angle is small for the large mass of a particle. The random
displacement of the segment end is of the order $\mu _{\mathrm{d}}\vartheta
_{\mathrm{dM}}=\sqrt{2d}$, i.e. of the same order as the segment width. It
is reasonable, because these two phenomena have the common source: the
space-time distortion $D$.

One should note that the space-time geometry influences the stochasticity of
particle motion nonlocally in the sense, that the form of the world function
(\ref{b3.3}) for values of $\sigma _{\mathrm{M}}<\frac{1}{2}\mu _{\mathrm{d}%
}^{2}$ is unessential for the motion stochasticity of the particle of the
mass $\mu _{\mathrm{d}}$.

Such a situation, when the world line of a free particle is stochastic in
the deterministic geometry, and this stochasticity depends on the particle
mass, seems to be rather exotic and incredible. But experiments show that
the motion of real particles of small mass is stochastic indeed, and this
stochasticity increases, when the particle mass decreases. From physical
viewpoint a theoretical foundation of the stochasticity is desirable, and
some researchers invent stochastic geometries, noncommutative geometries and
other exotic geometrical constructions, to obtain the quantum stochasticity.
But in the Riemannian space-time geometry the particle motion does not
depend on the particle mass, and in the framework of the Riemannian
space-time geometry it is difficult to explain the quantum stochasticity by
the space-time geometry properties. The distorted geometry $\mathcal{G}_{%
\mathrm{d}}$ explains freely the stochasticity and its dependence on the
particle mass. Besides, at proper choice of the distortion $d$ the
statistical description of stochastic $\mathcal{T}_{\mathrm{br}}$ leads to
the quantum description (in terms of the Schr\"{o}dinger equation) \cite{R91}%
. To do this, it is sufficient to set 
\begin{equation}
d=\frac{\hbar }{2bc}  \label{b3.12a}
\end{equation}%
where $\hbar $ is the quantum constant, $c$ is the speed of the light, and $%
b $ is some universal constant, connecting the geometrical mass $\mu $ with
the usual particle mass $m$ by means of the relation $m=b\mu $. In other
words, the distorted space-time geometry (\ref{b3.3}) is closer to the real
space-time geometry, than the Minkowski geometry $\mathcal{G}_{\mathrm{M}}$.

Further development of the statistical description of geometrical
stochasticity leads to a creation of the model conception of quantum
phenomena (MCQP), which relates to the conventional quantum theory
approximately in the same way as the statistical physics relates to the
axiomatic thermodynamics. MCQP is the well defined relativistic conception
with effective methods of investigation \cite{R03}, whereas the conventional
quantum theory is not well defined, because it uses incorrect space-time
geometry, whose incorrectness is compensated by additional hypotheses
(quantum principles). Besides, it has problems with application of the
nonrelativistic quantum mechanical technique to the description of
relativistic phenomena.

The geometry $\mathcal{G}_{\mathrm{d}}$, as well as the Minkowski geometry
are homogeneous geometries, because the world function $\sigma _{\mathrm{d}}$
is invariant with respect to all coordinate transformations, with respect to
which the world function $\sigma _{\mathrm{M}}$ is invariant. In this
connection the question arises, whether one could invent some axiomatics for 
$\mathcal{G}_{\mathrm{d}}$ and derive the geometry $\mathcal{G}_{\mathrm{d}}$
from this axiomatics by means of proper reasonings. Note that such an
axiomatics is to depend on the parameter $d$, because the world function $%
\sigma _{\mathrm{d}}$ depends on this parameter. If $d=0$, this axiomatics
is to coincide with the axiomatics of the Minkowski geometry $\mathcal{G}_{%
\mathrm{M}}$. If $d\neq 0$, this axiomatics cannot coincide with the
axiomatics of $\mathcal{G}_{\mathrm{M}}$, because some axioms of $\mathcal{G}%
_{\mathrm{M}}$ are not satisfied in this case. In general, the invention of
axiomatics, depending on the parameter $d$ and in the general case on the
distortion function $D$, seems to be a very difficult problem. Besides, why
invent the axiomatics? We had derived the axiomatics for the proper
Euclidean geometry, when we constructed it before. There is no necessity to
repeat this process any time, when we construct a new geometry. It is
sufficient to apply the deformation principle to the constructed Euclidean
geometry written $\sigma $-immanently. Application of the deformation
principle to the Euclidean geometry is a very simple and general procedure,
which is not restricted by continuity, convexity and other artificial
constraints, generated by our preconceived approach to the physical
geometry. (Bias of the approach is displayed in the antecedent supposition
on the one-dimensionality of any straight in any physical geometry).

Thus, we have seen that the nondegeneracy, as well as non-one-dimensionality
of the straight are properties of the real physical geometries. The proper
Euclidean geometry is a ground for all physical geometries, and it is a
degenerate geometry. Nevertheless, it is beyond reason to deny an existence
of nondegenerate physical geometries.

\section{Corollaries of the nondegenerate space-time \newline
geometry}

Possibility of the nondegenerate space-time geometry changes strongly the
existing conception of the microcosm space-time. A small correction to the
world function of the Minkowski space-time admits one to explain the
enigmatic quantum nature of the microcosm. The quantum principles as an
addition to the classical picture of the world appear to be not needed. The
quantum principles become to be corollary of the space-time model. The
microcosm space-time geometry changes radically. The universal transversal
length $\sqrt{d}$ (\ref{b3.12a}) appears as an attribute of the space-time.
The particle motion becomes primordially stochastic. One does not need to
search for the reason of stochasticity in the sense that the stochastic
motion is a natural motion of any particle, whereas the deterministic motion
is a motion of the particle of the extremely large mass.

In the classical mechanics (Newtonian or relativistic) the natural particle
motion is deterministic. If the particle moves stochastically, one should
search for the reason of the stochasticity. In the classical mechanics we
reduce the stochastic particle motion to the natural deterministic motion.
The stochastic motion is interpreted via the deterministic one. In the
nondegenerate space-time we must be able to perceive the stochastic motion
directly, without reducing it to the deterministic motion. It is a very
difficult problem, because one needs to create a new conception of
mechanics. Some ideas of such a conception of mechanics one can find in \cite%
{R2002,R004a}.

\section{Association problems instead of logical ones}

Logical problems are absent in T-geometry, because they are supposed to be
solved in the proper Euclidean geometry. Considering the space-time geometry
as a T-geometry, we meet the problem of separation between the geometry and
the dynamics.

The physical geometry is a science on mutual disposition of geometrical
objects. But any geometrical object is an abstraction. As a set of points,
any geometrical object does not exist in itself. In reality we may have some
substance, having a shape of the geometrical object. The set of points (the
geometrical object) is an abstraction of this fact. It is supposed, that we
can always to separate the shape of the substance from its contents.

Classical physics and classical dynamics accept, that the substance may have
any shape. Thus, it is accepted that any set of space points forms a
geometrical object, because any set of points can be filled by the
substance. This suggestion supposes the unlimited divisibility of the
substance and, hence, the unlimited divisibility of geometrical objects. All
this is valid for the usual space.

In the space-time we have another picture. In the classical physics the
particle is the simplest element of substance. Existence of a particle in
the space-time is described by the continuous tube, known as the particle
world tube. If the particle is point-like, the world tube degenerates into
the one-dimensional world line. The arbitrary geometrical object, i.e. an
arbitrary set of events (space-time points), cannot exist. If we suppose
that any substance consists of point-like particles, any geometrical object
is an arbitrary set infinite one-dimensional world lines. Only such a set of
events can be realized in the space-time as a geometrical object, because
such a set may be filled by the substance. The geometrical object, which can
be realized, i.e. filled with the substance, will be referred to as the
physical object. The world line of a point-like particle is supposed to be
indefinitely divisible, although this fact cannot be tested experimentally,
because the world line is always infinite or closed.

The one-dimensional curve (\ref{g1.1}) and its partial case -- the
one-dimensional straight form the basis of the Riemannian geometry as well
as of the classical dynamics. The property of the divisibility of the
continuous one-dimensional straight is formulated as follows. Let the
straight segment $\mathcal{T}_{\left[ P_{0}P_{1}\right] }$ between the
points $P_{0}$ and $P_{1}$%
\begin{eqnarray}
\mathcal{T}_{\left[ P_{0}P_{1}\right] } &=&\left\{ R|f_{\mathrm{s}%
P_{0}P_{1}}\left( R\right) =0\right\}  \label{h7.21} \\
f_{\mathrm{s}P_{0}P_{1}}\left( R\right) &=&\sqrt{2\sigma \left(
P_{0},R\right) }+\sqrt{2\sigma \left( R,P_{1}\right) }-\sqrt{2\sigma \left(
P_{0},P_{1}\right) }  \label{h7.21a}
\end{eqnarray}%
be divided into two parts $\mathcal{T}_{\left[ P_{0}Q\right] }$ and $%
\mathcal{T}_{\left[ QP_{1}\right] }$ by the point $Q$. If the straight is
continuous and one-dimensional, we obtain 
\begin{eqnarray}
\mathcal{T}_{\left[ P_{0}P_{1}\right] } &=&\mathcal{T}_{\left[ P_{0}Q\right]
}\cup \mathcal{T}_{\left[ QP_{1}\right] },\qquad \forall Q\in \mathcal{T}_{%
\left[ P_{0}P_{1}\right] }  \label{h7.22} \\
\mathcal{T}_{\left[ P_{0}Q\right] } &=&\left\{ R|f_{\mathrm{s}P_{0}Q}\left(
R\right) =0\right\} ,\qquad \mathcal{T}_{\left[ QP_{1}\right] }=\left\{ R|f_{%
\mathrm{s}QP_{1}}\left( R\right) =0\right\}  \nonumber
\end{eqnarray}%
The relations (\ref{h7.22}) are satisfied for any points $P_{0}$, $P_{2}$,
provided the geometry is Riemannian and all segments $\mathcal{T}_{\left[
P_{0}P_{1}\right] }$ are one-dimensional. The property (\ref{h7.22}) of the
one-dimensional straight (geodesic) will be referred to as the divisibility
property.

Repeating division of the segment $\mathcal{T}_{\left[ P_{0}P_{1}\right] }$
many times, we obtain%
\begin{eqnarray}
\mathcal{T}_{\left[ P_{0}P_{1}\right] } &=&\dbigcup\limits_{i=0}^{i=N}%
\mathcal{T}_{\left[ Q_{i}Q_{i+1}\right] },\qquad Q_{0}=P_{0},\qquad
Q_{N+1}=P_{1}  \label{h7.23} \\
Q_{k} &\in &\mathcal{T}_{\left[ Q_{k-1}Q_{k+1}\right] },\qquad \forall
Q_{k}\in \mathcal{T}_{\left[ P_{0}P_{1}\right] },\qquad k=1,2,...N
\label{h7.24}
\end{eqnarray}%
The divisibility admits one to divide the segment $\mathcal{T}_{\left[
P_{0}P_{1}\right] }$ into infinitesimal segments and reestablish the primary
segment $\mathcal{T}_{\left[ P_{0}P_{1}\right] }$ by its infinitesimal
parts. It is important that any infinitesimal segment $\mathcal{T}_{\left[
Q_{i}Q_{i+1}\right] }$ is an element of the world line. It turns into
several elements of the world line after further division.

The unlimited divisibility of a one-dimensional continuous curve is a ground
for the infinitesimal analysis, created by Newton and Leibniz. As far as any
geometrical objects of the Riemannian geometry and objects of classical
mechanics (trajectories in the phase space and world lines in the
space-time) may be considered as consisting of infinitesimal straight line
segments, the infinitesimal analysis appears to be the mathematical tool of
geometry and physics. The possibility of the world line restoration by its
infinitesimal elements is a necessary property of the unlimited
divisibility. Simultaneously it is a necessary condition of the
infinitesimal analysis applicability.

In the nondegenerate geometry, where the straight is not one-dimensional, in
general, we have, 
\begin{equation}
\mathcal{T}_{\left[ P_{0}Q\right] }\nsubseteqq \mathcal{T}_{\left[ P_{0}P_{1}%
\right] },\qquad \mathcal{T}_{\left[ QP_{1}\right] }\nsubseteqq \mathcal{T}_{%
\left[ P_{0}P_{1}\right] },\qquad Q\in \mathcal{T}_{\left[ P_{0}P_{1}\right]
}  \label{h7.25}
\end{equation}%
It means that we can divide the segment $\mathcal{T}_{\left[ P_{0}P_{1}%
\right] }$ into parts, but we cannot reestablish it by its parts, in
general, because the relation (\ref{h7.22}) does not take place.

Of course, we can divide the segment $\mathcal{T}_{\left[ P_{0}Q\right] }$
into two parts, cutting it at the point $Q\neq P_{0},P_{1}$. For instance, 
\begin{eqnarray*}
\mathcal{T}_{\mathrm{c}\left[ P_{0}Q\right] P_{1}} &=&\left\{ R|f_{\mathrm{s}%
P_{0}P_{1}}\left( R\right) =0\wedge \sqrt{2\sigma \left( P_{0},R\right) }%
\leq \sqrt{2\sigma \left( P_{0},Q\right) }\right\} ,\qquad Q\in \mathcal{T}_{%
\left[ P_{0}P_{1}\right] } \\
\mathcal{T}_{\mathrm{c}\left[ P_{1}Q\right] P_{0}} &=&\left\{ R|f_{\mathrm{s}%
P_{0}P_{1}}\left( R\right) =0\wedge \sqrt{2\sigma \left( P_{1},R\right) }%
\leq \sqrt{2\sigma \left( P_{1},Q\right) }\right\} ,\qquad Q\in \mathcal{T}_{%
\left[ P_{0}P_{1}\right] }
\end{eqnarray*}%
\begin{equation}
\mathcal{T}_{\left[ P_{0}P_{1}\right] }=\mathcal{T}_{\mathrm{c}\left[ P_{0}Q%
\right] P_{1}}\cup \mathcal{T}_{\mathrm{c}\left[ P_{1}Q\right] P_{0}}
\label{h7.28}
\end{equation}%
But the sets $\mathcal{T}_{\mathrm{c}\left[ P_{0}Q\right] P_{1}}$ and $%
\mathcal{T}_{\mathrm{c}\left[ P_{1}Q\right] P_{0}}$ do not coincide with the
straight line segments $\mathcal{T}_{\left[ P_{0}Q\right] }$ $\mathcal{T}_{%
\left[ P_{1}Q\right] }$. In particular, the set $\mathcal{T}_{\mathrm{c}%
\left[ P_{0}Q\right] P_{1}}$ depends on the external point $P_{1}\notin 
\mathcal{T}_{\mathrm{c}\left[ P_{0}Q\right] P_{1}}$. It means that we may
not consider the sets $\mathcal{T}_{\mathrm{c}\left[ P_{0}Q\right] P_{1}}$
and $\mathcal{T}_{\mathrm{c}\left[ P_{1}Q\right] P_{0}}$ as a physical
objects, i.e. as the set of points which can be filled by the substance.
Besides, the fact that $\mathcal{T}_{\mathrm{c}\left[ P_{0}Q\right] P_{1}}$
depends on the external point $P_{1}\notin \mathcal{T}_{\mathrm{c}\left[
P_{0}Q\right] P_{1}}$ means that the set $\mathcal{T}_{\mathrm{c}\left[
P_{0}Q\right] P_{1}}$ is a part of $\mathcal{T}_{\left[ P_{0}P_{1}\right] }$%
, whereas $\mathcal{T}_{\left[ P_{0}Q\right] }$ may be considered to be a
self-contained object, but not a part of other geometrical object. We shall
formulate the difference between $\mathcal{T}_{\mathrm{c}\left[ P_{0}Q\right]
P_{1}}$ and $\mathcal{T}_{\left[ P_{0}Q\right] }$ as follows. The set of
points $\mathcal{T}_{\left[ P_{0}Q\right] }$ forms a physical object,
whereas the set of points $\mathcal{T}_{\mathrm{c}\left[ P_{0}Q\right]
P_{1}} $ is not a physical object. The set of points $\mathcal{T}_{\mathrm{c}%
\left[ P_{0}Q\right] P_{1}}$ is a part of a physical object $\mathcal{T}_{%
\left[ P_{0}P_{1}\right] }$. Here we meet a strange situation, when a part
of a physical object may be not a physical object. In other words, the
physical object has a finite size and cannot be divided into physical
objects of infinitesimal size.

Lack of unlimited divisibility (finite divisibility) of the simplest
geometrical object (the straight segment) leads to the finite divisibility
of more complicated geometrical objects. The finite divisibility is
associated in a way with the "quantum nature" of microcosm. The "quantum
nature" of the microcosm is rather cloudy concept. It means a description of
physical phenomena by means of the quantum mechanics formalism. The "quantum
nature" means something connected rather with the dynamics, than with the
geometry. The finite divisibility is a property of the nondegenerate
space-time geometry. It is mainly a geometrical property, although there is
some reference to the properties of the matter. But it is important, that
the finite divisibility is such a property, which can be investigated by the
geometrical methods, as soon as the world function of the space-time is
known. In any case the concept of the finite divisibility is less cloudy,
than the concept of "quantum nature".

\section{The boundary between physics and geometry}

Any physical phenomenon takes place in the space-time. Existence of the
space-time is a common circumstance for all physical phenomena. Description
of each physical phenomenon has two sides: (1) the side, common for all
physical phenomenon, known as space-time geometry, (2) the side, specific
for each physical phenomenon, known as dynamics. It means that some
characteristics of the physical phenomenon are considered as geometrical
characteristics, whereas other ones are considered as dynamical
characteristics. The boundary between geometry and dynamics is not constant.
This boundary varies, as the physics developed.

In the time of Isaac Newton the geometry was the most developed branch of
mathematics. Geometrical methods of investigation were dominating. The
famous Newtonian book \cite{N1687} was written mainly in terms of geometry.
Creation of the infinitesimal analysis by Newton and Leibniz, and its
application in mechanics changed the situation. Analytical methods of
description and investigation became dominating. Analytical methods
penetrated into the geometry, and the analytical geometry arose. Further
development of the geometry and, in particular, generalization of the
Euclidean geometry was produced by the analytical methods in the framework
of the differential geometry. The methods of the infinitesimal analysis were
effective at description of only such physical processes, which could be
divided into a set of local processes.

On the other hand, the role of the space-time geometry in the description of
the dynamic systems increased. Some important physical concepts, considered
at first as purely dynamical, appeared to have a geometrical origin. The
energy-momentum conservation laws, considered at first as properties of
dynamical systems, appeared to be generated by the space-time geometry. The
quantum phenomena, considered in the XXth century as specific dynamic
phenomena, are looking now as generated by the space-time properties. In
particular, the quantum constant $\hbar $ appears to be a parameter of the
space-time geometry. The particle mass appears to be a geometrical
characteristic. Now one may not speak on the special quantum nature of
physical phenomena in microcosm. All they are explained freely by means of
the properties of the space-time geometry. As our knowledge on the microcosm
physical phenomena develops, the role of the space-time geometry increases,
and the geometry -- dynamics boundary is shifted towards the dynamics.

In general, it seems rather reasonable, that all general physical laws are
to have the geometrical origin, because the space-time is the only common
circumstance, which takes place in all physical phenomena. For instance, the
conservation law of the electric charge and multiplicity of any electrical
charge to the elementary one are to have the geometrical origin, because
these phenomena are universal. For the present these phenomena are not
always connected with the space-time geometry.

In the nondegenerate space-time geometry there is no ground for the
infinitesimal analysis, which supposes unlimited divisibility of the
physical objects and a possibility of the primary physical object
restoration by its parts. Note that the world tube divisibility cannot be
tested experimentally. It is such a supposition which is tested by its
corollaries.

In such a situation we may not imagine that we can divide the world tube of
a real particle into parts. We are forced to suppose that the division of
the world tube into elementary parts exists objectively, and such a division
is not a result of our method of the world tube description. The different
divisions of the world tube correspond to the world tubes of different
particles. The world tube of a real particle is a broken tube, i.e. a chain
of the connected finite segments $\mathcal{T}_{\left[ P_{k}P_{k+1}\right] }$%
, whose length is proportional to the particle mass, whereas the vector $%
\mathbf{P}_{k}\mathbf{P}_{k+1}$ describes the particle momentum. In the
nondegenerate space-time geometry one cannot abstract from the size of the
link, which is connected with the particle mass. As a result the particle
mass become to be a geometrical characteristic of a particle. In other
words, the particle mass is geometrized. At the conventional consideration
the particle mass is its dynamical characteristic. Thus, in the
nondegenerate space-time geometry the boundary between geometry and dynamics
is shifted towards the dynamics.

Any link has a finite size, and the conventional mathematical technique of
the infinitesimal analysis does not work in the nondegenerate space-time
geometry. The problem of adequate mathematical description of geometrical
objects in the nondegenerate space-time geometry appears.

\section{Interplay of geometric and analytical methods}

Construction of T-geometry by means of the local analytical methods appeared
to be ineffective. Application of analytical methods to the geometry needs
an introduction of continuous coordinate systems, which restricts strongly
the list of possible geometries. Application of nonlocal finite methods of
description (the deformation principle) appears to be more effective at the
geometry generalization. In the nondegenerate space-time with tubes instead
of straights an application of the differential geometry appears to be
ineffective at the construction of local characteristics of the space-time
geometry.

The fact is that the methods of differential geometry are connected closely
with concept of the one-dimensional curve as a primary geometrical object of
the geometry. In the geometry, where the one-dimensional curve is not a
physical object, the local analytical methods do not work effectively. One
does not develop the nonlocal geometrical methods, which are based on the
description of the whole geometrical object, but not on its infinitesimal
elements. As a result we were forced to use conventional analytical methods
in the quantum mechanics substantiation, founded on the geometry and
statistics, In this case the distortion of the space-time geometry was
considered as a correction to the Minkowski space-time geometry.

Let us consider a simple example, which shows that not only description in
terms of infinitesimal analysis is possible. We consider solution of the
Klein-Gordon equation in the Riemannian space-time. In the Minkowski
space-time this equation has the form 
\begin{equation}
\left( g^{ik}\partial _{i}\partial _{k}+\frac{m^{2}}{\hbar ^{2}c^{2}}\right)
\psi =0,\qquad g^{ik}=\text{diag}\left\{ c^{-2},-1,-1,-1\right\}
\label{h7.1}
\end{equation}

The causal Green function for the equation (\ref{h7.1}) has the form \cite%
{BS57}

\begin{eqnarray}
D^{c}\left( x-x^{\prime }\right) &=&\frac{1}{4\pi }\delta \left( 2\sigma
\right) -\frac{mc}{8\pi \hbar \sqrt{2\sigma }}\theta \left( \sigma \right) %
\left[ J_{1}\left( \frac{mc\sqrt{2\sigma }}{\hbar }\right) -iN_{1}\left( 
\frac{mc\sqrt{2\sigma }}{\hbar }\right) \right]  \nonumber \\
&&+i\frac{mc}{4\pi ^{2}\hbar \sqrt{-2\sigma }}\theta \left( -\sigma \right)
K_{1}\left( -\frac{mc\sqrt{2\sigma }}{\hbar }\right) ,  \label{h7.2} \\
\qquad \theta \left( x\right) &=&\left\{ 
\begin{array}{c}
1,\text{ if }x>0 \\ 
0,\text{ if }x<0%
\end{array}%
\right.  \nonumber
\end{eqnarray}%
where $\sigma $ is the world function 
\begin{equation}
\sigma =\sigma \left( x,x^{\prime }\right) =\frac{1}{2}\left( c^{2}\left(
x^{0}-x^{\prime 0}\right) ^{2}-\left( \mathbf{x}-\mathbf{x}^{\prime }\right)
^{2}\right)  \label{h7.3}
\end{equation}%
and $J_{1}$, $N_{1}$ and $K_{1}$ are the first order cylindric functions
(respectively of Bessel, Neumann and Hankel).

Obtaining of the causal Green function (\ref{h7.2}) is equivalent to a
solution of the Klein-Gordon equation (\ref{h7.1}).

In the Riemannian space-time the equation (\ref{h7.1}) turns into the
equation 
\begin{equation}
\left( g^{ik}\nabla _{i}\nabla _{k}+\frac{m^{2}}{\hbar ^{2}c^{2}}\right)
\psi =0,  \label{h7.4}
\end{equation}
where now $g^{ik}$ is the metric tensor of the Riemannian space-time, and $%
\nabla _{k}$ is the covariant derivative in this space-time.

If we consider the process, which is described by the Klein-Gordon equation
in the curved space-time, we may use two different methods:

\begin{enumerate}
\item To solve the equation (\ref{h7.4}).

\item To take the expression (\ref{h7.2}) for the causal Green function,
where $\sigma $ is the world function of the Riemannian space-time. It may
be determined as the symmetric solution of the Jacobi-Hamilton equation 
\begin{equation}
g^{ik}\left( \partial _{i}\sigma \right) \partial _{k}\sigma =2\sigma
,\qquad \sigma \left( x,x^{\prime }\right) =\sigma \left( x^{\prime
},x\right)  \label{h7.5}
\end{equation}
\end{enumerate}

The second method is simpler in the sense that the equation (\ref{h7.5}) is
the first order differential equation.

We are not sure that both methods lead to the same result. We do not discuss
here, which of two methods is valid. We should like to pay attention to the
following circumstance. The second (integral) method may be applied, \textit{%
at least formally}, in the curved space-time and in the distorted space-time
(\ref{b3.3}). In the last case the world function in (\ref{h7.2}) is taken
in the form (\ref{b3.3}). The first (differential) method cannot be applied
in the distorted space-time (\ref{b3.3}) even formally, because it is not
clear how to take into account the space-time distortion.

\section{Problems of the T-geometry perception}

Although the T-geometry construction is very simple, it appears to be rather
difficult for perception by mathematicians \cite{R2005}. Its perception is
especially difficult for professional geometers, which know and apply the
mathematical formalism of the differential geometry. The fact is that the
T-geometry cannot be constructed by means of the formalism of the
differential geometry and that of the infinitesimal analysis, which is a
foundation of the differential geometry.

The Riemannian geometry is constructed as a generalization of the proper
Euclidean geometry, presented in the form of differential geometry. The
generalization is produced on the basis of such a geometrical object as the
Euclidean straight. Such properties of the Euclidean straight as continuity
and one-dimensionality are conserved at the generalization, whereas such
properties as vanishing curvature and vanishing torsion of the Euclidean
straight may be violated. In the Euclidean geometry, as well as in the
Riemannian geometry the concept of one-dimensional continuous curve is the
primary object of the geometry. The concept of the curve is primary in the
sense that any geometrical object may be considered as consisting of a set
of curves. In turn any curve may be considered as consisting of infinite set
of infinitesimal segments of a curve. Infinitesimal segments of a curve can
be considered as infinitesimal segments of the straight. Thus, any
geometrical object can be constructed of infinitesimal segments of a curve
(straight), which are the primary objects of the Riemannian geometry.

On the other hand, the infinitesimal analysis has been invented by Newton
and Leibniz for description of functions

\begin{equation}
y=f\left( x\right)  \label{h8.1}
\end{equation}%
Such a function describes a curve on the plane $\left( x,y\right) $. Any
geometrical object can be constructed of infinitesimal segments of the curve
(\ref{h8.1}). It means that the infinitesimal analysis is applicable for
description of the differential geometry and, in particular, for description
of the Riemannian geometry. Applicability of the infinitesimal analysis to
problems of mechanics and physics is connected with the fact that problems
of these sciences can be described in terms of curves, or geometrical
objects, consisting of curves.

The T-geometry is the generalization of the Euclidean geometry, founded on
the $\sigma $-immanence property of the Euclidean geometry, which can be
presented in terms of the world function $\sigma _{\mathrm{E}}$. The $\sigma 
$-immanence is the property of the whole Euclidean geometry, whereas the
straight is only a geometrical object of the Euclidean geometry. The
straight is very important geometrical object, but it is only one
geometrical object among many others. Generalization made on the basis of
the property of the whole Euclidean geometry is more preferable, than the
generalization made on the basis of properties of one object. Thus, from the
common viewpoint the T-geometry is more preferable, than the Riemannian
geometry. The one-dimensional continuous curve is not a primary geometrical
object of the T-geometry. It can be introduced as a secondary geometrical
object, i.e. as a geometrical object, obtained by intersection of several
primary geometrical objects, for instance, (\ref{c2.1}), (\ref{c2.2}). In
some special cases the first order tube may degenerate into the
one-dimensional straight. In this case we may use the means of the
infinitesimal analysis. In the general case the one-dimensional curve is not
a primary geometrical object and an application of the infinitesimal
analysis appears to be ineffective.

\textit{Remark.} In the Riemannian geometry the one-dimensional curve is the
primary geometrical object of the geometry. The surface may be constructed
of curves. This fact is reasonable and customary. In T-geometry the
one-dimensional curve is a secondary (derivative) geometrical object, which
can be constructed of some primary objects (surfaces) by means of their
intersection. It seems to be incredible, that in T-geometry the curve is
constructed of surfaces. I remember very well, that I could not perceive
this fact for a long time.

The geometry can be described not only in terms of infinitesimal quantities
by means of the infinitesimal analysis. It can be described also in terms of
finite geometrical quantities. Unfortunately, the particle dynamics can be
described only in terms of infinitesimal quantities by methods of the
infinitesimal analysis. In contemporary physics the geometrical methods are
not used in the particle dynamics. The particle dynamics is described in
terms of one-dimensional curves (trajectories in the phase space, world
lines in the space-time). The infinitesimal analysis describes these
geometrical objects very well. The problem of the geometrical methods
application did not arise in the contemporary particle dynamics.

In the nondegenerate space-time T-geometry (distorted space-time (\ref{b3.3}%
)) the particle world tube is not a one-dimensional curve. It is a chain
consisting of finite non-one-dimensional links. In the simplest case any
link is the straight segment $\mathcal{T}_{\left[ P_{k}P_{k+1}\right] }$,
whose length is proportional to the particle mass $m$. Quantum effects are
connected with the thickness $\sqrt{d}$ and the length $\left\vert \mathcal{T%
}_{\left[ P_{k}P_{k+1}\right] }\right\vert $ of the link, and we may not to
tend the length $\left\vert \mathcal{T}_{\left[ P_{k}P_{k+1}\right]
}\right\vert $ to zero. Hence, we cannot reduce the problem of the particle
dynamics directly to the description in terms of the one-dimensional curve
and solve it by methods of the infinitesimal analysis. Unfortunately, in the
particle dynamics we have no methods of description except for infinitesimal
analysis. We are forced to describe the chain as a one-dimensional line. The
link shape (thickness and length) was taken into account as some stochastic
agent. This agent makes the one-dimensional world line to be stochastic.
Statistical description of this stochastic motion leads to the quantum
description in terms of the Schr\"{o}dinger equation. But such a result we
obtain in the simplest case of structure-less particle.

Analysis of the free Dirac particle, (i.e. the dynamic system $\mathcal{S}_{%
\mathrm{D}}$, described by the Dirac equation) shows \cite{R004}, that the
Dirac particle has a complex structure. This structure may be interpreted
either dynamically, or geometrically. According to the dynamic
interpretation the Dirac particle is a rotator, i.e. two constituents
rotating around their common center of inertia. According to the geometrical
interpretation the world tube of the Dirac particle is the chain of links $%
\mathcal{T}_{\left[ P_{i}Q_{i}P_{i+1}\right] }$, $i=0,\pm 1,...$ Any link $%
\mathcal{T}_{\left[ P_{i}Q_{i}P_{i+1}\right] }$ is a triangle with vertices
at the points $P_{i},Q_{i},P_{i+1}$. In the case of the dynamical
interpretation we have a confinement problem, i.e. we are to explain, what
forces coupled the constituents of the Dirac particle. In the case of the
geometrical interpretation we are not to explain anything. If in the
simplest case the link may be a segment $\mathcal{T}_{\left[ P_{i}P_{i+1}%
\right] }$ of the first order tube, then why cannot it be a segment $%
\mathcal{T}_{\left[ P_{i}Q_{i}P_{i+1}\right] }$ of the second order tube, or
even a segment $\mathcal{T}_{\left[ P_{i}Q_{i}R_{i}P_{i+1}\right] }$ of the
third order tube? The geometrical interpretation seems to be preferable,
because it looks more natural, and besides, it does not contain the
confinement problem. The geometrical approach looks especially attractive in
the case of the segment $\mathcal{T}_{\left[ P_{i}Q_{i}R_{i}P_{i+1}\right] }$%
, where three vertices $P_{i},Q_{i},R_{i}$ are associated with three quarks
inside the composite particle. The confinement problem is absent at the
geometrical approach.

Although the model of the Dirac particle is not quite perfect, because
internal degrees of freedom are described non-relativistically \cite{R04},
but this model is the best model of the relativistic particle, and one
should not ignore results of its investigation. Besides, the nonrelativistic
character of this model can be corrected \cite{R04}.

Calculation of the elementary particles mass spectrum is considered now as
the main problem of the elementary particles theory. The problem of the
elementary particles mass spectrum is considered conventionally as a
dynamical problem, which reminds the problem of the atom electromagnetic
emanation spectrum. One tries to solve this problem, searching for an
appropriate dynamic system.

In the nondegenerate space-time geometry the particle mass is geometrized,
and one should expect that the problem of the elementary particles mass
spectrum is a geometrical problem, but not a dynamical one. In other words,
we should search for appropriate structure of links, constituting the
particle world tubes, but not for dynamic systems imitating this structure.
Investigation of the dynamic system is based mainly on properties of groups,
connected with the structure of the link. Such indirect investigation is
more complicated, than a direct investigation of the geometrical structure
of the link. Besides, considering the dynamical system, we are to take into
account that the dynamical interaction propagates with the speed less, than
the speed of the light. If we consider a structure with the purely
geometrical couplings, we are free of this constraint, because the geometry
is more fundamental part of the description, than the dynamics. The
principles of relativity impose constraints only on dynamic systems, but not
on the geometrical objects. We may solve the confinement problem without
introduction of gluons.

The distorted space-time with particles, described by non-one-dimensional
world tubes, sets the problem of adequate mathematical formalism. This
formalism is to deal with the finite (but not infinitesimal) quantities.
Besides, it must be oriented to different geometrical objects (whereas the
infinitesimal analysis is oriented to an infinitesimal segment of
one-dimensional curve). Construction of such a mathematical technique is a
very complicated problem. Until such a mathematical technique is not
created, we are forced to reduce the geometrical problems to the dynamical
ones and to use the infinitesimal analysis.

\section{Concluding remarks}

The T-geometry is a very simple generalization of the Euclidean geometry.
Construction of T-geometry expands the class of physical geometries
appropriate for description of the space-time. Among homogeneous isotropic
Riemannian geometries there is only one geometry appropriate for the
space-time description. This is the Minkowski geometry. In the class of
T-geometries there is a set of homogeneous isotropic geometries. Any
homogeneous isotropic T-geometry is labelled by the distortion function $%
D\left( \sigma _{\mathrm{M}}\right) $, describing the shape and thickness of
the timelike straight segments. One may treat this circumstance as an
introduction of the transverse universal length. The free particle motion is
primordially deterministic only in the Minkowski space-time geometry with $%
D\left( \sigma _{\mathrm{M}}\right) =0$. In all other homogeneous, isotropic
T-geometries the free particle motion is primordially stochastic.

When in the beginning of XXth century it has been discovered, that the free
motion of the small mass particle is primordially stochastic, one should to
choose one of T-geometries with $D\left( \sigma _{\mathrm{M}}\right) \neq 0$
as a space-time geometry. Unfortunately, in that time neither the
non-degenerate T-geometries, nor the nondegeneracy property were not known.
Researchers were forced to use the Minkowski space-time geometry everywhere,
including microcosm. To obtain the corollaries of the stochastic particle
motion and to explain physical phenomena of microcosm, they are forced to
introduce \textit{additional hypotheses}, known as quantum principles.
Because of lack of any alternative the quantum principles are accorded wide
recognition, and now most researchers speak on the quantum origin of
microcosm.

When the T-geometry had been constructed, it became clear, that the physical
phenomena of microcosm can be freely explained by a true choice of the
space-time geometry. There is no necessity to introduce the quantum
principles and to speak about the quantum nature of the microcosm physical
phenomena. The quantum nature of the microcosm prescribes to quantize all
physical fields, including metrical fields, which describe the space-time
properties. In the nondegenerate space-time geometry we have the distortion
and the finite divisibility instead of the quantum nature, and there is no
necessity to quantize the metrical fields. The electromagnetic and
gravitational fields are metrical fields, because in the 5-dimensional
Klein-Kaluza geometry they describe the properties of the space-time. One
failed to quantize the gravitational field, and there is no problem with it.
But the electromagnetic field has been quantized successfully, and at this
point we have disagreement between the two approaches. There are no
experimental evidence in favour of the quantization necessity. Experiments
show that the electromagnetic field is emitted and absorbed in the form of
quanta. This fact may be explained by the properties of the emitting or
absorbing atom. But does the electromagnetic field \textit{exist} in the
form of quanta? We do not know experiments, which could answer positively
this question. Besides, the dynamic equations for the electromagnetic and
gravitational fields do not contain the quantum constant. Thus, there are
doubts in the necessity of the electromagnetic field quantization.

Consideration of the nondegenerate T-geometry as a space-time geometry is
the third essential revision of our space-time conception.

The first essential revision of the space-time conception was the transition
from the Newtonian model of the space-time with two invariants to the
Minkowski space-time geometry with one invariant. The second essential
revision of the space-time conception was the transition from homogeneous
space-time geometry to the nonhomogeneous space-time geometry, where
nonhomogeneity is generated by the substance distribution. The third
revision of the space-time model is founded on existence of a new class of
physical geometries, having such unknown unusual properties as the
nondegeneracy and lack of the unlimited divisibility. These new properties
of the space-time geometry concerns mainly microcosm. As any revision, the
new conception of the space-time is difficult for perception. It is valid
especially for those researchers, who believe that the geometry is a
totality of coordinates, metric tensor, covariant derivatives etc.

Application of the T-geometry to the microcosm physics generates problems,
connected with the further geometrization of physics. Interpretation of new
properties of the T-geometry and a construction of an adequate mathematical
formalism are the main problems of the new conception of the space-time.


\begin{thebibliography}{99}
\bibitem{K37} F. Klein, \textit{Vorlesungen \"{u}ber die Entwicklung die
Mathematik im 19. Jahrhundert} teil 1, Berlin, Springer 1926.

\bibitem{S60} J. L.~Synge, \textit{Relativity: The General Theory},
North-Holland, Amsterdam, 1960.

\bibitem{R90} Yu. A.~Rylov, Extremal properties of Synge's world function
and discrete geometry. \textit{J. Math. Phys.} \textbf{31}, 2876-2890,
(1990).

\bibitem{R01} Yu.A. Rylov, Geometry without topology as a new conception of
geometry.\textit{\ Int. Jour. Mat. \& Mat. Sci.} \textbf{30}, iss. 12,
733-760, (2002), (available at http://arXiv.org/abs/math.MG/0103002).

\bibitem{R002} Yu. A. Rylov, Asymmetric geometry without topology as a
geometry of microcosm. in \textit{Geometrical and Topological Ideas in
Modern Physics (Proceedings of the XXV workshop on the fundamental problems
of high energy physics and field theory). }Protvino, June 25-28, 2002, pp.
154-190. \newline
Another version "Asymmetric nondegenerate geometry". (Available at
http://arXiv.org/abs/math.MG/0205061 ).

\bibitem{R91} Yu. A.~Rylov, Non-Riemannian model of space-time responsible
for quantum effects. \textit{J. Math. Phys.} \textbf{32}, 2092-2098, (1991).

\bibitem{H30} D. Hilbert, \textit{Grundlagen der Geometrie}. 7 Auflage,
B.G.Teubner, Leipzig, Berlin, 1930.

\bibitem{M28} K.~Menger, Untersuchen \"{u}ber allgemeine Metrik, \textit{%
Mathematische Annalen,} \textbf{100}, 75-113, (1928).

\bibitem{B53} L. M.~Blumenthal, \textit{Theory and Applications of Distance
Geometry}, Oxford, Clarendon Press, 1953.

\bibitem{R02} Yu. A. Rylov, Deformation principle and problem of parallelism
in geometry and physics.\textit{\ (}In preparation, available at
http://arXiv.org/abs/ math.GM /0210413)

\bibitem{R03} Yu. A. Rylov, Model conception of quantum phenomena: logical
structure and investigation methods. (In preparation, available at
http://arXiv.org/abs /physics/0310050, v2).

\bibitem{N1687} I. Newton, \textit{Philosophiae naturalis principia
mathematica}, 1687.

\bibitem{BS57} N.N. Bogoliubov and D.V. Shirkov, \textit{Introduction to
theory of quantum fields.} Moscow, 1957, GITTL. (in Russian) p.432.

\bibitem{R2005} Yu. A. Rylov, New crisis in geometry? (Available at
http://arXiv.org/ abs/mat.GM/0503261).

\bibitem{R2004} Yu. A. Rylov, Deformation principle as a foundations of
physical geometry and its application to the space-time geometry.
((Available at http://arXiv.org/abs/physics/0411103)

\bibitem{R2002} Yu. A. Rylov, Dynamics of stochastic systems and
peculiarities of measurements in them (Available at
http://arXiv.org/abs/physics/0210003).

\bibitem{R004a} Yu. A. Rylov, What object does the wave function describe?
(Available at http://arXiv.org/abs/physics/0405117).

\bibitem{R004} Yu. A. Rylov, Is the Dirac particle composite? (Available at
http://arXiv.org /abs/physics /0410045).

\bibitem{R04} Yu. A. Rylov, Is the Dirac particle completely relativistic?
(Available at http://arXiv.org/abs/physics/0412032).
\end{thebibliography}
\end{document}